\documentclass[aps,prd,showpacs]{revtex4}
\usepackage{graphicx,epsfig,xcolor}
\usepackage{dcolumn,bm}
\def\a{\alpha}
\def\b{\beta}

\def\g{\gamma}

\def\s{\sigma}
\def\S{\Sigma}
\def\o{\omega}
\def\O{\Omega}
\def\L{\Lambda}
\def\p{\prime}

\def\GpiSS{g_{\pi\S_c\S_c}}
\def\GpiXX{g_{\pi\Xi_c\Xi_c}}
\def\GpiXXp{g_{\pi\Xi_c^\p\Xi_c^\p}}
\def\GetaXX{g_{\eta\Xi_c\Xi_c}}
\def\GetaLL{g_{\eta\L_c\L_c}}
\def\GetaSS{g_{\eta\S_c\S_c}}
\def\GetaOO{g_{\eta\O_c\O_c}}
\def\GetaXXp{g_{\eta\Xi_c^\p\Xi_c^\p}}
\def\GsigLL{g_{\s\L_c\L_c}}
\def\GsigXX{g_{\s\Xi_c\Xi_c}}
\def\GsigXXp{g_{\s\Xi_c^\p\Xi_c^\p}}
\def\GsigSS{g_{\s\S_c\S_c}}
\def\GsigOO{g_{\s\O_c\O_c}}
\def\GomLL{g_{\o\L_c\L_c}}
\def\FomLL{f_{\o\L_c\L_c}}
\def\GomXX{g_{\o\Xi_c\Xi_c}}
\def\FomXX{f_{\o\Xi_c\Xi_c}}
\def\GomXXp{g_{\o\Xi_c^\p\Xi_c^\p}}
\def\FomXXp{f_{\o\Xi_c^\p\Xi_c^\p}}
\def\GomSS{g_{\o\S_c\S_c}}
\def\FomSS{f_{\o\S_c\S_c}}
\def\GrhoXX{g_{\rho\Xi_c\Xi_c}}
\def\FrhoXX{f_{\rho\Xi_c\Xi_c}}
\def\GrhoXXp{g_{\rho\Xi_c^\p\Xi_c^\p}}
\def\FrhoXXp{f_{\rho\Xi_c^\p\Xi_c^\p}}
\def\GrhoSS{g_{\rho\S_c\S_c}}
\def\FrhoSS{f_{\rho\S_c\S_c}}
\def\GphiXX{g_{\phi\Xi_c\Xi_c}}
\def\FphiXX{f_{\phi\Xi_c\Xi_c}}
\def\GphiXXp{g_{\phi\Xi_c^\p\Xi_c^\p}}
\def\FphiXXp{f_{\phi\Xi_c^\p\Xi_c^\p}}
\def\GphiOO{g_{\phi\O_c\O_c}}
\def\FphiOO{f_{\phi\O_c\O_c}}

\def\GpiNN{g_{\pi NN}}
\def\GetaNN{g_{\eta NN}}
\def\GsigNN{g_{\s NN}}
\def\GrhoNN{g_{\rho NN}}
\def\FrhoNN{f_{\rho NN}}
\def\GomNN{g_{\o NN}}
\def\FomNN{f_{\o NN}}

\def\GpiQQ{g_{\pi qq}}
\def\GetaQQ{g_{\eta qq}}
\def\GrhoQQ{g_{\rho qq}}
\def\GomQQ{g_{\o qq}}
\def\GsigQQ{g_{\s qq}}
\def\GphiQQ{g_{\phi qq}}

\def\NO{\nonumber}
\def\kev{\mathrm{~keV}}
\def\mev{\mathrm{~MeV}}
\def\gev{\mathrm{~GeV}}

\def\fm{\mathrm{~fm}}

\begin{document}

\title{Possible Deuteron-like Molecular States Composed of Heavy Baryons}
\author{Ning Lee}\email{leening@pku.edu.cn}
\author{Zhi-Gang Luo}\email{cglow@pku.edu.cn}
\author{Xiao-Lin Chen}\email{chenxl@pku.edu.cn}
\author{Shi-Lin Zhu}\email{zhusl@pku.edu.cn}
\affiliation{Department of Physics
and State Key Laboratory of Nuclear Physics and Technology\\
Peking University, Beijing 100871, China}
\date{\today}

\begin{abstract}

We perform a systematic study of the possible loosely bound states
composed of two charmed baryons or a charmed baryon and an
anti-charmed baryon within the framework of the one boson exchange
(OBE) model. We consider not only the $\pi$ exchange but also the
$\eta$, $\rho$, $\o$, $\phi$ and $\s$ exchanges. The $S-D$ mixing
effects for the spin-triplets are also taken into account. With
the derived effective potentials, we calculate the binding
energies and root-mean-square (RMS) radii for the systems
$\L_c\L_c(\bar{\L}_c)$, $\Xi_c\Xi_c(\bar{\Xi}_c)$,
$\S_c\S_c(\bar{\S}_c)$, $\Xi_c^\p\Xi_c^\p(\bar{\Xi}_c^\p)$ and
$\O_c\O_c(\bar{\O}_c)$. Our numerical results indicate that: (1)
the H-dibaryon-like state $\L_c\L_c$ does not exist; (2) there may
exist four loosely bound deuteron-like states $\Xi_c\Xi_c$ and
$\Xi_c^\p\Xi_c^\p$ with small binding energies and large RMS radii.
.

\end{abstract}


\maketitle

\section{Introduction\label{intro}}

Many so-called ``XYZ'' charmonium-like states such as $X(3872)$,
$X(4350)$ and $Y(3940)$ have been observed by Belle, CDF, D0 and
BaBar collaborations~\cite{Belle,BaBar,CDF,D0} during the past few
years. Despite the similar production mechanism, some of these
structures do not easily fit into the conventional charmonium
spectrum, which implies other interpretations such as hybrid
mesons, heavy meson molecular states etc. might be responsible for
these new states~\cite{Brambilla:2010cs}\cite{Swanson2006}.

A natural idea is that some of the ``XYZ'' states near two heavy
meson threshold may be bound states of a pair of heavy meson and
anti-heavy meson.  Actually, Rujula {\em et al.} applied this idea
to explain $\psi(4040)$ as a P-wave $D^*\bar{D}^*$ bound resonance
in the 1970s~\cite{Rujula77}. Tornqvist performed an intensive
study of the possible deuteron-like two-charm-meson bound states
with the one-pion-exchange (OPE) potential model in
Ref.~\cite{Torq}. Recently, motivated by the controversy over the
nature of $X(3872)$ and $Z(4430)$, some authors proposed $X(3872)$
might be a $D\bar{D}^*$ bound
state~\cite{Swan04,Wong04,Close2004,Voloshin2004,Thomas2008}. Our
group have studied the possible molecular structures composed of a
pair of heavy mesons in the framework of the One-Boson-Exchange
(OBE) model systematically~\cite{LiuXLiuYR,ZhugrpDD}.  There are
also many interesting investigations of other hadron
clusters~\cite{Ding,LiuX,Liu2009,Ping:2000dx,Liu:2011xc,qiao}.

The boson exchange  models are very successful to describe nuclear
force~\cite{Mach87,Mach01,Rijken}. Especially the deuteron is a
loosely bound state of proton and neutron, which may be regarded
as a hadronic molecular state.  One may wonder whether a pair of
heavy baryons can form a deuteron-like bound state through the
light meson exchange mechanism. On the other hand, the large
masses of the heavy baryons reduce the kinetic of the systems,
which makes it easier to form bound states. Such a system is
approximately non-relativistic. Therefore, it is very interesting
to study whether the OBE interactions are strong enough to bind
the two heavy baryons (dibaryon) or a heavy baryon and an
anti-baryon (baryonium).


A heavy charmed baryon contains a charm quark and two light
quarks. The two light quarks form a diquark. Heavy charmed baryons
can be categorized by the flavor wave function of the diquark,
which form a symmetric $6$ or an antisymmetric $\bar{3}$
representation. For the ground heavy baryon, the spin of the
diquark is either $0$ or $1$, and the spin of the baryon is either
$1/2$ or $3/2$. The product of the diquark flavor and spin wave
functions of the ground charmed baryon must be symmetric and
correlate with each other. Thus the spin of the sextet diquark is
$1$ while the spin of the anti-triplet diquark is $0$.

The ground charmed baryons are grouped into one antitrpilet with
spin-1/2 and two sextets with spin-1/2 and spin-3/2 respectively.
These multiplets are usually denoted as $B_{\bar{3}}$, $B_{6}$ and
$B_{6}^*$ in literature \cite{Yan}. In the present work, we study
the charmed dibaryon and baryonium systems, i.e.
$\L_c\L_c(\bar{\L}_c)$, $\Xi_c\Xi_c(\bar{\Xi}_c)$,
$\S_c\S_c(\bar{\S}_c)$, $\Xi^\p_c\Xi^\p_c(\bar{\Xi}_c^\p)$ and
$\O_c\O_c(\bar{\O}_c)$. Other configurations will be explored in a
future work. We first derive the effective potentials of these
systems. Then we calculate the binding energies and
root-mean-square (RMS) radii to determine which system
might be a loosely bound molecular state.

This work is organized as follows. We present the formalism in
section \ref{form}. In section \ref{const}, we discuss the
extraction of the coupling constants between the heavy baryons and
light mesons and give the numerical results in Section
\ref{numer}. The last section is a brief summary. Some useful
formula and figures are listed in appendix.

\section{Formalism\label{form}}

In this section we will construct the wave functions and derive
the effective potentials.

\subsection{Wave Functions}

\begin{figure}[htb]
	  \hfill
	  \begin{minipage}[b]{0.5\textwidth}
	  \centering
	  \includegraphics[width=0.6\textwidth]{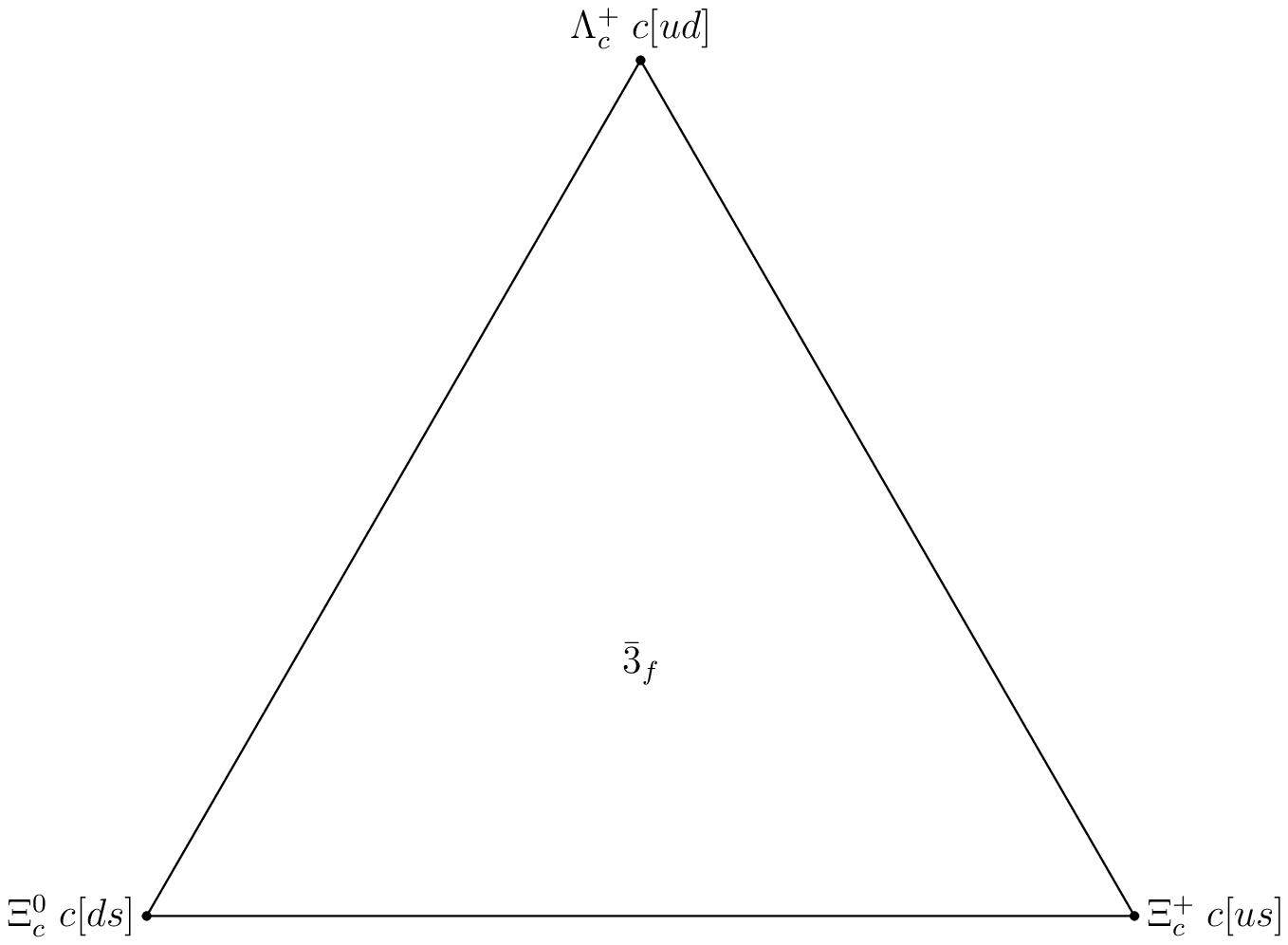}\\ (a) antitriplet
	\end{minipage}%
	\hfill
	  \begin{minipage}[b]{0.5\textwidth}
	  \centering
	  \includegraphics[width=0.6\textwidth]{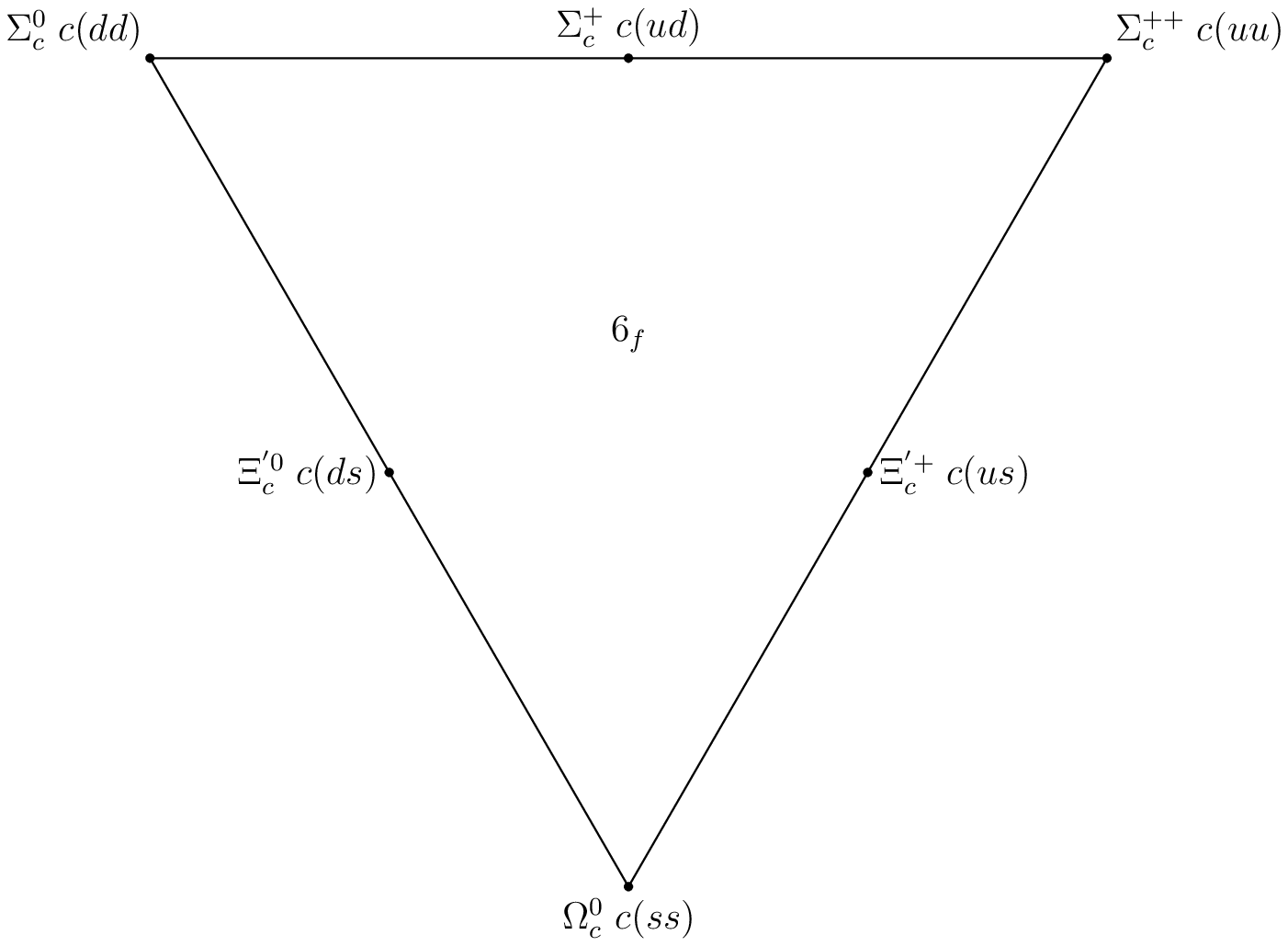}\\ (b) sextet
		\end{minipage}\hspace*{\fill}
  \caption{The antitriplet and sextet. Here the brackets and parentheses
represent antisymmetrization and symmetrization of the light
	quarks respectively.}
  \label{Fig:Multi} 
\end{figure}

As illustrated in Fig.~\ref{Fig:Multi}, the states $\L_c^+$,
$\Xi_c^+$ and $\Xi_c^0$ belong to the antitriplet $B_{\bar{3}}$
while $\S_c^{++}$, $\S_c^{+}$, $\S_c^0$, $\Xi_c^{\p+}$,
$\Xi_c^{\p0}$ and $\O_c^0$ are in sextet $B_6$.  Among them,
$\L_c^+$ and $\O_c^0$ are isoscalars; $\{\Xi_c^+,\Xi_c^0\}$ and
$\{\Xi_c^{\p+},\Xi_c^{\p0}\}$ are isospin spinnors;
$\{\S_c^{++},\S_c^+,\S_c^0\}$ is an isovector. We denote these
states $\L_c$, $\Xi_c$, $\S_c$, $\Xi_c^\p$ and $\O_c$.

The wave function of a dibaryon is the product of its isospin,
spatial and spin wave functions,
\begin{eqnarray}
\Psi_{hh}^{[I,2S+1]}\sim\Psi_{hh}^I\otimes\Psi_{hh}^L\otimes\Psi_{hh}^S\label{Eq:HHfunc}.
\end{eqnarray}
We consider the isospin function $\Psi_{hh}^I$ first. The isospin
of $\L_c$ is $0$, so $\L_c\L_c$ has isospin $I=0$ and
$\Psi_{\L_c\L_c}^{I=0}=\L_c^+\L_c^+$,
which is symmetric. For $\Xi_c\Xi_c$, the isospin is $I=0$ or $1$,
and their corresponding wave functions are antisymmetric and
symmetric respectively.  $\S_c\S_c$ has isospin $0$, $1$ or $2$.
Their flavor wave functions can be constructed using
Clebsch-Gordan coefficients. $\Xi_c^\p\Xi_c^\p$ is the same as
$\Xi_c\Xi_c$.  The isospin of the $\O_c\O_c$ is $0$. Because
strong interactions conserve isospin symmetry, the effective
potentials do not depend on the third components of the isospin.
For example,
it is adequate to take the isospin function $\Xi_c^+\Xi_c^+$ with $I_3=1$ when we derive the
effective potential for $\Psi_{\Xi_c\Xi_c}^{I=1}$, though the wave
function $\frac{1}{\sqrt{2}}(\Xi_c^+\Xi_c^0+\Xi_c^0\Xi_c^+)$
indeed gives the same result. In the following, we show the
relevant isospin functions used in our calculation,
\begin{eqnarray}
\Psi_{\L_c\L_c}^{I=0} &=& \L_c^+\L_c^+\\
\Psi_{\Xi_c\Xi_c}^{I=0} &=&
\frac{1}{\sqrt{2}}\left(\Xi_c^+\Xi_c^0-\Xi_c^0\Xi_c^+\right)\NO\\
\Psi_{\Xi_c\Xi_c}^{I=1} &=& \Xi_c^+\Xi_c^+\\
\Psi_{\S_c\S_c}^{I=0} &=& \frac{1}{\sqrt{3}}\left(\S_c^{++}\S_c^0-\S_c^+\S_c^++\S_c^0\S_c^{++}\right)\NO\\
\Psi_{\S_c\S_c}^{I=1} &=&\frac{1}{\sqrt{2}}\left(\S_c^{++}\S_c^+-\S_c^+\S_c^{++}\right)\NO\\
\Psi_{\S_c\S_c}^{I=2} &=& \S_c^{++}\S_c^{++}\\
\Psi_{\Xi_c^\p\Xi_c^\p}^{I=0} &=&
\frac{1}{\sqrt{2}}\left(\Xi_c^{\p+}\Xi_c^{\p0}-\Xi_c^{\p0}\Xi_c^{\p+}\right)\NO\\
\Psi_{\Xi_c^\p\Xi_c^\p}^{I=1} &=& \Xi_c^{\p+}\Xi_c^{\p+}\\
\Psi_{\O_c\O_c}^{I=0} &=& \O_c^0\O_c^0.
\end{eqnarray}

We are mainly interested in the ground states of dibaryons and
baryonia where the spatial wave functions of these states are
symmetric. The tensor force in the effective potentials mixes the
$S$ and $D$ waves. Thus a physical ground state is actually a
superposition of the $S$ and $D$ waves. This mixture fortunately does
not affect the symmetries of the spatial wave functions.
As a mater of fact, for a dibaryon with a specific total spin $\bm
J$, we must add the spins of its components to form $\bm S$ first
and then couple $\bm S$ and the relative orbit angular momentum
$\bm L$ together to get $\bm J=\bm L+\bm S$. This $L-S$ coupling
scheme leads to six $S$ and $D$ wave states: $^1S_0$, $^3S_1$,
$^1D_2$, $^3D_1$, $^3D_2$ and $^3D_3$. But the tenser force only
mixes states with the same $S$ and $J$. In our case we must deal
with the $^3S_1$-$^3D_1$ mixing. After stripping off the isospin
function, the mixed wave function is
\begin{eqnarray}
|\psi\rangle=R_S(r)|^3S_1\rangle+R_D(r)|^3D_1\rangle,
\end{eqnarray}
which will lead to  coupled channel Schr\"odinger equations for
the radial functions $R_S(r)$ and $R_D(r)$. In short, for the
spatial wave functions, we will discuss the ground states in
$^1S_0$ and $^3S_1$, and the latter mixes with $^3D_1$.

Finally, we point out that the $I$ and $S$ of states in
Eq.~(\ref{Eq:HHfunc}) can not be combined arbitrarily because the
generalized identity principle constricts the wave functions to be
antisymmetric. It turns out that the survived compositions are
$\Psi_{\L_c\L_c}^{[0,1]}$, $\Psi_{\Xi_c\Xi_c}^{[0,3]}$,
$\Psi_{\Xi_c\Xi_c}^{[1,1]}$,
$\Psi_{\S_c\S_c}^{[0,1]}$,$\Psi_{\S_c\S_c}^{[1,3]}$,$\Psi_{\S_c\S_c}^{[2,1]}$,$\Psi_{\Xi_c^\p\Xi_c^\p}^{[0,3]}$,
$\Psi_{\Xi_c^\p\Xi_c^\p}^{[1,1]}$, and $\Psi_{\O_c\O_c}^{[0,1]}$.
For baryonia, there is no constraint on the wave functions. So we
need take into account more states. The wave functions of baryonia
can be constructed in a similar way. However, we can use the so-called ``G-Parity
rule'' to derive the effective potentials for baryonia directly
from the corresponding potentials for dibaryons, and it is no need
discussing them here now.

\subsection{Lagrangians}

We introduce notations
\begin{eqnarray}
\L_c=\L_c^+,\quad \Xi_c=\left(\begin{array}{c}
						\Xi_c^+ \\ \Xi_c^0
						\end{array}\right), \quad \bm\S_c=\left\{\frac{1}{\sqrt{2}}(-\S_c^{++}+\S_c^0),\frac{i}{\sqrt{2}}(-\S_c^{++}-\S_c^0),\S_c^+\right\},
\quad \Xi_c^\p=\left( \begin{array}{c} \Xi_c^{\p+}\\ \Xi_c^{\p0}
\end{array}
\right), \quad \O_c=\O_c^0
\end{eqnarray}
to represent the corresponding baryon fields. The long range
interactions are provided by the $\pi$ and $\eta$ meson exchanges:
\begin{eqnarray}
\mathcal{L}_{\pi} &=& \GpiXX\bar{\Xi}_c i\g_5\bm\tau\Xi_c\cdot\bm\pi
+\GpiSS(-i)\bar{\bm\S}_c i\g_5\times\bm\S_c\cdot\bm\pi + \GpiXXp \bar{\Xi}^\p_c
i\g_5\bm\tau\Xi^\p_c\cdot\bm\pi\\
\mathcal{L}_\eta &=& \GetaLL\bar{\L}_ci\g_5\L_c\eta +
\GetaXX\bar{\Xi}_ci\g_5\Xi_c\eta \NO\\ && + \GetaSS\bar{\bm\S}_c\cdot
i\g_5\bm\S_c\eta + \GetaXXp\bar{\Xi}_c^\p i\g_5\Xi_c^\p\eta +
\GetaOO\bar{\O}_ci\g_5\O_c\eta,
\end{eqnarray}
where $\GpiXX$, $\GpiSS$, $\GetaOO$ etc. are the coupling
constants. $\bm \tau=\{\tau_1,\tau_2,\tau_3\}$ are the Pauli
matrices, and
$\bm\pi=\{\frac{1}{\sqrt{2}}(\pi^++\pi^-),\frac{i}{\sqrt{2}}(\pi^+-\pi^-),\pi^0\}$
are the $\pi$ fields.  The vector meson exchange
Lagrangians read
\begin{eqnarray}
\mathcal{L}_{\rho} &=& \GrhoXX\bar{\Xi}_c\g_\mu\bm\tau\Xi_c\cdot\bm\rho^\mu +
\frac{\FrhoXX}{2m_{\Xi_c}}\bar{\Xi}_c\s_{\mu\nu}\bm\tau\Xi_c\cdot\partial^\mu\bm\rho^{\nu}
\NO\\ && + \GrhoSS(-i)\bar{\bm\S}_c\g_\mu\times\bm\S_c\cdot\bm\rho^\mu +
\frac{\FrhoSS}{2m_{\S_c}}(-i)\bar{\bm\S}_c\s_{\mu\nu}\times\bm\S_c\cdot\partial^\mu\bm\rho^\nu
\NO\\ && + \GrhoXXp\bar{\Xi}_c^\p\g_\mu\bm\tau\Xi_c^\p\cdot\bm\rho^\mu +
\frac{\FrhoXXp}{2m_{\Xi^\p_c}}\bar{\Xi}_c^\p\s_{\mu\nu}\bm\tau\Xi_c^\p\cdot\partial^\mu\bm\rho^\nu\\
\mathcal{L}_\o &=& \GomLL\bar{\L}_c\g_\mu\L_c\o^\mu +
\frac{\FomLL}{2m_{\L_c}}\bar{\L}_c\s_{\mu\nu}\L_c\partial^\mu\o^\nu \NO\\ &&
+\GomXX\bar{\Xi}_c\g_\mu\Xi_c\o^\mu +
\frac{\FomXX}{2m_{\Xi_c}}\bar{\Xi}_c\s_{\mu\nu}\Xi_c\partial^\mu\o^\nu
\NO\\ && +\GomSS\bar{\bm\S}_c\g_\mu\cdot\bm\S_c\o^\mu +
\frac{\FomSS}{2m_{\S_c}}\bar{\bm\S}_c\s_{\mu\nu}\cdot\bm\S_c\partial^\mu\o^\nu
\NO\\ && +\GomXXp\bar{\Xi}_c^\p\g_\mu\Xi_c^\p\o^\mu +
\frac{\FomXXp}{2m_{\Xi_c^\p}}\bar{\Xi}_c^\p\s_{\mu\nu}\Xi_c^\p\partial^\mu\o^\nu\\
\mathcal{L}_\phi &=& \GphiXX\bar{\Xi}_c\g_\mu\Xi_c\phi^\mu +
\frac{\FphiXX}{2m_{\Xi_c}}\bar{\Xi}_c\s_{\mu\nu}\Xi_c\partial^\mu\phi^\nu
\NO\\ && +\GphiXXp\bar{\Xi}_c^\p\g_\mu\Xi_c^\p\phi^\mu
+\frac{\FphiXXp}{2m_{\Xi_c^\p}}\bar{\Xi}_c^\p\s_{\mu\nu}\Xi_c^\p\partial^\mu\phi^\nu \NO\\ &&
+\GphiOO\bar{\O}_c\g_\mu\O_c\phi^\mu +
\frac{\FphiOO}{2m_{\O_c}}\bar{\O}_c\s_{\mu\nu}\O_c\partial^\mu\phi^\nu,
\end{eqnarray}
with
$\bm\rho=\{\frac{1}{\sqrt{2}}(\rho^++\rho^-),\frac{i}{\sqrt{2}}(\rho^+-\rho^-),\rho^0\}$.
The $\s$ exchange Lagrangian is
\begin{eqnarray}
\mathcal{L}_\s &=& \GsigLL\bar{\L}_c\L_c\s + \GsigXX\bar{\Xi}_c\Xi_c\s +
\GsigSS\bar{\bm\S}_c\cdot\bm\S_c\s \NO\\ && + \GsigXXp\bar{\Xi}^\p_c\Xi^\p_c\s
+\GsigOO\bar{\O}_c\O_c\s.
\end{eqnarray}
There are thirty-three unknown coupling constants in the above Lagrangains,
which will be determined in Sec.~\ref{const}.

\subsection{Effective Potentials}

\begin{figure}[htb]
\begin{center}
\hfill\includegraphics[width=0.2\textwidth]{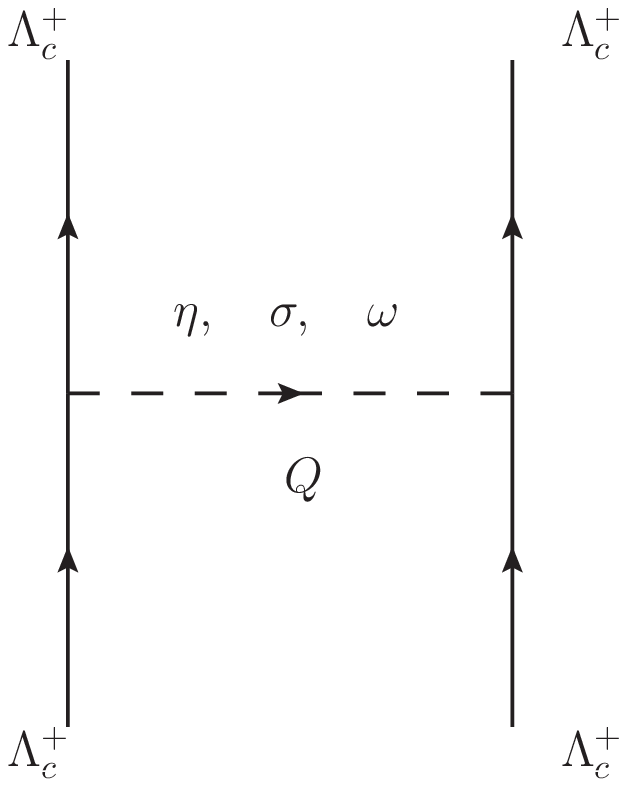}%
\hfill\includegraphics[width=0.2\textwidth]{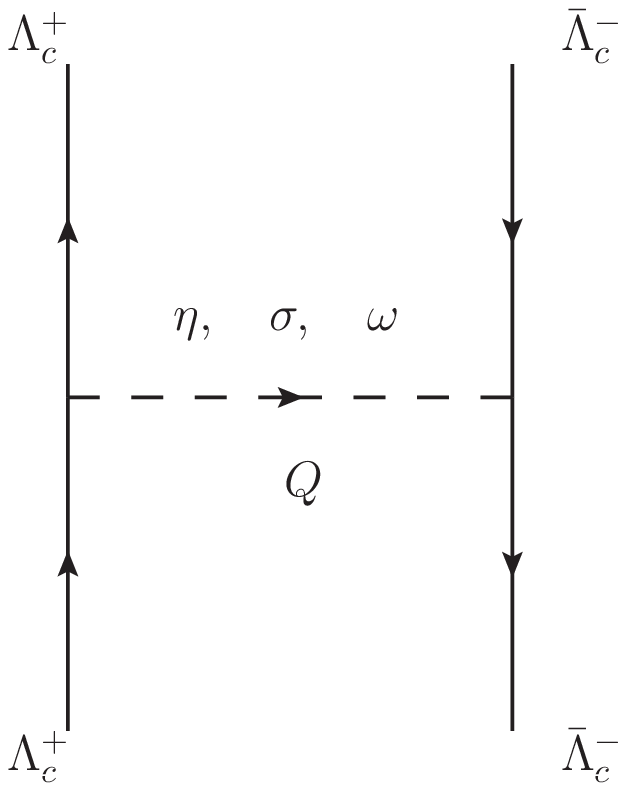}\hspace*{\fill}
\caption{Scattering processes of $\L_c\L_c\to\L_c\L_c$ and
  $\L_c\bar{\L}_c\to\L_c\bar{\L}_c$, $Q$s are the transformed four momenta.}
\label{Fig:scatt}
\end{center}
\end{figure}

To obtain the effective potentials, we calculate the $T$ matrices of
the scattering processes such as Fig.~\ref{Fig:scatt} in momentum
space. Expanding the $T$ matrices with external momenta to the
leading order, one gets \cite{Barnes:1999hs}
\begin{eqnarray}
V(\bm r)=\frac{1}{(2\pi)^3}\int d^3{q} e^{-i\bm Q\cdot\bm r} T(\bm Q)
\mathcal{F}(\bm Q)^2,
\end{eqnarray}
where $\mathcal{F}(\bm Q)$ is the form factor, with which the
divergency in the above integral is controlled, and the
non-point-like hadronic structures attached to each vertex are
roughly taken into account. Here we choose the monopole form
factor
\begin{eqnarray}
\mathcal{F}(\bm Q)=\frac{\L^2-m^2}{\L^2-Q^2}
\end{eqnarray}
with $Q=\{Q_0,\bm Q\}$ and the cutoff $\L$.

Generally speaking, a potential derived from the scattering $T$
matrix consists of the central term, spin-spin interaction term,
orbit-spin interaction term and tenser force term, i.e.,
\begin{eqnarray}
V(\bm r)=V_C(r)+V_{SS}(r)\bm\s_1\cdot\bm\s_2+V_{LS}(r)\bm L\cdot\bm S+ V_T(r)S_{12}(\hat{\bm r}),
\end{eqnarray}
where $S_{12}(\hat{\bm r})$ is the tensor force operator,
$S_{12}(\hat{\bm
  r})=3(\bm\s_1\cdot\hat{\bm r})(\bm\s_2\cdot\hat{\bm
  r})-\bm\s_1\cdot\bm\s_2$.  The effective potential of a specific channel,
for example $\L_c\L_c\to\L_c\L_c$ shown in Fig.~\ref{Fig:scatt},
may contain contributions from the pseudoscalar, vector and scalar
meson exchanges. We need work them out one by one and add them.
The potentials with the stripped isospin factors from the
pseudoscalar, vector and scalar ($\s$ here) meson exchange are
\begin{eqnarray}
V^a(\bm
r;\a,h)&=&V^a_{SS}(r;\a,h)\bm\s_1\cdot\bm\s_2+V^a_T(r;\a,h)S_{12}(\hat{\bm
  r}),\NO\\ V^b(\bm
r;\b,h)&=&V^b_C(r;\b,h)+V^b_{SS}(r;\b,h)\bm\s_1\cdot\bm\s_2+V^b_{LS}(r;\b,h)\bm
L\cdot\bm S+V^b_T(r;\b,h)S_{12}(\hat{\bm r}),\NO\\ V^c(\bm
r;\s,h)&=&V^c_C(r;\s,h)+V^c_{LS}(r;\s,h)\bm L\cdot\bm S\label{EQ:Vabc},
\end{eqnarray}
where $\a=\pi,\eta$, $\b=\rho,\o,\phi$ and
\begin{eqnarray}
V^a_{SS}(r;\a,h)&=&-\frac{g_{\a
	hh}^2}{4\pi}\frac{m_\a^3}{12m_h^2}H_1(\L,m_\a,r),\NO\\ V^a_T(r;\a,h)&=&\frac{g_{\a
	hh}^2}{4\pi}\frac{m_\a^3}{12m_h^2}H_3(\L,m_\a,r),\NO\\ V^b_C(r;\b,h)&=&\frac{m_\b}{4\pi}\left[g_{\b
	hh}^2H_0(\L,m_\b,r)-(g_{\b hh}^2+4g_{\b hh}f_{\b
	hh})\frac{m_\s^2}{8m_h^2}H_1(\L,m_\b,r)\right],\NO\\ V^b_{SS}(r;\b,h)&=&-\frac{1}{4\pi}(g_{\b
  hh}+f_{\b
  hh})^2\frac{m_\b^3}{6m_h^2}H_1(\L,m_\b,r),\NO\\ V^b_{LS}(r;\b,h)&=&-\frac{1}{4\pi}(3g_{\b
  hh}^2+4g_{\b hh}f_{\b
  hh})\frac{m_\b^3}{2m_h^2}H_2(\L,m_\b,r),\NO\\ V^b_T(r;\b,h)&=&-\frac{1}{4\pi}(g_{\b
  hh}+f_{\b
  hh})^2\frac{m_\b^3}{12m_h^2}H_3(\L,m_\b,r),\NO\\ V^c_C(r;\s,h)&=&-m_\s\frac{g_{\s
	hh}^2}{4\pi}\left[H_0(\L,m_\s,r)+\frac{m_\s^2}{8m_h^2}H_1(\L,m_\s,r)\right],\NO\\ V^c_{LS}(r;\s,h)&=&-m_\s\frac{g_{\s
	hh}^2}{4\pi}\frac{m_\s^2}{2m_h^2}H_2(\L,m_\s,r).
\end{eqnarray}
The definitions of functions $H_0$, $H_1$, $H_2$ and $H_3$ are
given in the appendix. From Eq.~(\ref{EQ:Vabc}), one can see the
tensor force terms and spin-spin terms are from the pseudoscalar
and vector meson exchanges while the central and obit-spin terms
are from the vector and scalar meson exchanges. Finally the
effective potential of the state $hh$ is
\begin{eqnarray}
V_{hh}(\bm r)&=& \sum_\a \mathcal{C}^a_\a V^a(\bm r;\a,h) + \sum_\b \mathcal{C}^b_\b V^b(\bm
r;\b,h) + \mathcal{C}^c_\s V^c(\bm r;\s,h)\NO\\ &=& \left\{\sum_\b \mathcal{C}^b_\b
V^b_C(r;\b,h)+\mathcal{C}^c_\s V^c_C(r;\s,h)\right\}+\left\{\sum_\a \mathcal{C}^a_\a
V^a_{SS}(r;\a,h)+\sum_\b \mathcal{C}^b_\b
V^b_{SS}(r;\b,h)\right\}\bm\s_1\cdot\bm\s_2\NO\\ &&+\left\{\sum_\b \mathcal{C}^b_\b
V^b_{LS}(r;\b,h)+\mathcal{C}^c_\s V^c_{LS}(r;\b,h)\right\}\bm L\cdot\bm S+\left\{\sum_\a
\mathcal{C}^a_\a V^a_T(r;\a,h)+\sum_\b \mathcal{C}^b_\b V^b_T(r;\b,h)\right\}S_{12}(\hat{\bm
  r})\label{EQ:Vhh},\NO\\
\end{eqnarray}
where $\mathcal{C}^a_\a$, $\mathcal{C}^b_\b$ and
$\mathcal{C}^c_\s$ are the isospin factors, which are listed in
Table~\ref{Tab:isoFct}.

\begin{center}
\begin{table}[htb]
\begin{tabular}{c|c|cc|ccc|cc|c}
\hline\hline
~   &~~$\Lambda_c\Lambda_c[\bar{\Lambda}_c]$~~&\multicolumn{2}{c|}{~~$\Xi_c\Xi_c[\bar{\Xi}_c]$~~}&
\multicolumn{3}{c|}{~~$\Sigma_c\Sigma_c[\bar{\Sigma}_c]$~~}&\multicolumn{2}{c|}{~~$\Xi_c^{'}\Xi_c^{'}
[\bar{\Xi}_c^{'}]$~~}&~~$\Omega_c\Omega_c[\bar{\Omega}_c]$ \\
\hline
			  I   &  0   &   0   & 1 &   0   & 1 & 2 &  0   & 1  &  0   \\
\hline
$\mathcal{C}_{\pi}^a$   &   &  &	 &  -2[2]& -1[1]& 1[-1]& -3[3]& 1[-1]& \\
$\mathcal{C}_{\eta}^a$  &   &  &	 &   1[1]&  1[1]&  1[1]&  1[1]&  1[1]&  1[1] \\
$\mathcal{C}_{\rho}^b$  &   & -3[-3]&  1[1]& -2[-2]&-1[-1]&  1[1]&-3[-3]&  1[1]&  1[1] \\
$\mathcal{C}_{\omega}^b$&1[-1]  &1[-1]  & 1[-1]&  1[-1]& 1[-1]& 1[-1]& 1[-1]& 1[-1]&	   \\
$\mathcal{C}_{\phi}^b$  &   &  1[-1]& 1[-1]&  &  &   & 1[-1]& 1[-1]& 1[-1] \\
$\mathcal{C}_{\sigma}^c$&1[1]   & 1[1]  &  1[1]&   1[1]&  1[1]&  1[1]&  1[1]&  1[1]&	   \\
\hline\hline
\end{tabular}
\caption{Isospin factors. The values in  brackets for
baryonia are derived by the ``G-Parity rule''.}\label{Tab:isoFct}
\end{table}
\end{center}

Given the effective potential $V_{hh}$, the potential for
$h\bar{h}$, $V_{h\bar{h}}$, can be obtained using the ``G-Parity
rule'', which states that the amplitude (or the effective
potential) of the process $A\bar{A}\to A\bar{A}$ with one light
meson exchange is related to that of the process $AA\to AA$ by
multiplying the latter by a factor $(-)^{I_G}$, where $(-)^{I_G}$
is the G-Parity of the exchanged light meson~\cite{IGrule}. The expression of
$V_{h\bar{h}}$ is the same as Eq.~(\ref{EQ:Vhh}) but with $V^a(\bm
r;\a,h)$, $V^b(\bm r;\b,h)$ and $V^c(\bm r;\s,h)$ replaced by
$V^a(\bm r;\a,\bar{h})$, $V^b(\bm r;\b,\bar{h})$ and $V^c(\bm
r;\s,\bar{h})$ respectively.
\begin{eqnarray}
V^a(\bm r;\a,\bar{h})&=&(-)^{I_G[\a]}V^a(\bm r;\a,h),\NO\\
V^b(\bm r;\b,\bar{h})&=&(-)^{I_G[\b]}V^b(\bm r;\b,h),\NO\\
V^c(\bm r;\s,\bar{h})&=&(-)^{I_G[\s]}V^c(\bm r;\s,h).
\end{eqnarray}
For example,
\begin{eqnarray}
V^a(\bm r;\o,\bar{\L}_c)&=&(-1)V^a(\bm r;\o,\L_c),
\end{eqnarray}
since the G-Parity of $\o$ is negative.  In other words, we can
still use the right hand side of Eq.~(\ref{EQ:Vhh}) to calculate
$V_{h\bar{h}}$ but with the redefined isospin factors
\begin{eqnarray}
\mathcal{C}^a_\a\to(-)^{I_G[\a]}\mathcal{C}^a_\a,\;\;
\mathcal{C}^b_\b\to(-)^{I_G[\b]}\mathcal{C}^b_\b,\;\;
\mathcal{C}^c_\s\to(-)^{I_G[\s]}\mathcal{C}^c_\s,
\end{eqnarray}
which are listed in Table~\ref{Tab:isoFct} too.

The treatments of operators $\bm\s_1\cdot\bm\s_2$, $\bm L\cdot\bm S$ and $S_{12}(\hat{\bm r})$
are straightforward. For $^1S_0$,
\begin{eqnarray}
\bm\s_1\cdot\bm\s_2=-3,\;\; \bm L\cdot\bm S=0,\;\; S_{12}(\hat{\bm r})=0,
\end{eqnarray}
which lead to  single channel Shr\"odinger equations. But for
$^3S_1$, because of mixing with $^3D_1$, the above operators
should be represented in the
$\left\{|^3S_1\rangle,|^3D_1\rangle\right\}$ space, i.e.,
\begin{eqnarray}
\bm\s_1\cdot\bm\s_2=\left(\begin{array}{cc}
1 & 0\\ 0 & 1
\end{array}\right),\;\; \bm L\cdot\bm S=\left(\begin{array}{cc}
0 & 0\\ 0 & -3
\end{array}\right),\;\; S_{12}(\hat{\bm r})=\left(\begin{array}{cc}
0 & \sqrt{8}\\ \sqrt{8} & -2
\end{array}\right).
\end{eqnarray}
These representations lead to the coupled channel Shr\"odinger
equations.

\section{Coupling Constants\label{const}}

It is difficult to extract the coupling constants in the
Lagrangians experimentally. We may estimate them using the
well-known nucleon-meson coupling constants as inputs with the
help of the quark model. The details of this method are provided in
Ref.~\cite{Riska2001}. The one-boson exchange Lagrangian at the quark
level is
\begin{eqnarray}
\mathcal{L}_q &=& \GpiQQ\left(\bar{u}i\g_5u\pi^0 - \bar{d}i\g_5d\pi^0\right)
\NO\\ && + \GetaQQ\left(\bar{u}i\g_5u\eta + \bar{d}i\g_5d\eta -
2\bar{s}i\g_5s\eta\right) \NO\\ && +\GrhoQQ\left(\bar{u}\g_\mu u\rho^{0\mu} -
\bar{d}\g_\mu d\rho^{0\mu}\right)  \NO\\ && +\GomQQ\left(\bar{u}\g_\mu
u\o^\mu + \bar{d}\g_\mu d\o^\mu\right) + \GphiQQ \bar{s}\g_\mu s\phi^\mu
\NO\\ && +\GsigQQ\left(\bar{u}u\s+\bar{d}d\s+\bar{s}s\s\right) + \cdots,
\end{eqnarray}
where $\GpiQQ$, $\GetaQQ$, $\ldots$, $\GsigQQ$ are the coupling
constants of the light mesons and quarks. The vector meson terms
in this Lagrangian do not contain the anomalous magnetic moment
part because the constituent quarks are treated as point-like
particles. At the hadronic level, for instance, the
nucleon-nucleon-meson interaction Lagrangian reads
\begin{eqnarray}
\mathcal{L}_{NN} &=& \GpiNN\bar{N}i\g_5\bm\tau N\cdot\bm\pi + \GetaNN\bar{N}i\g_5
N\eta \NO\\ && +\GrhoNN\bar{N}\g_\mu\bm\tau N\cdot\bm\rho^\mu +
\frac{\FrhoNN}{2m_N}\bar{N}\s_{\mu\nu}\bm\tau N\cdot\partial^\mu\bm\rho^\nu
\NO\\ && + \GomNN\bar{N}\g_\mu N\o^\mu +
\frac{\FomNN}{2m_N}\bar{N}\s_{\mu\nu}N\partial^\mu\o^\nu \NO\\ && + \GsigNN\bar{N}N\s,
\end{eqnarray}
where $\GpiNN$, $\GetaNN$, $\ldots$, $\GsigNN$ are the coupling
constants. We  calculate the matrix elements for a specific
process both at quark and hadronic levels and then match them. In this way, we get
relations between the two sets of coupling constants,
\begin{eqnarray}
\GpiNN=\frac{5}{3}\GpiQQ\frac{m_N}{m_q},\;\;
\GetaNN=\GetaQQ\frac{m_N}{m_q},\NO\\ \GomNN=3\GomQQ,\;\;
\frac{\GomNN+\FomNN}{m_N}=\frac{\GomQQ}{m_q},\NO\\ \GrhoNN=\GrhoQQ,\;\;
\frac{\GrhoNN+\FrhoNN}{m_N}=\frac{5}{3}\frac{\GrhoQQ}{m_q},\NO\\ \GsigNN=3\GsigQQ.\label{coupNN}
\end{eqnarray}
From these relations, we can see that $\GomNN$ and $\FomNN$ are
not independent. So are $\GrhoNN$ and $\FrhoNN$. The constituent
quark mass is about one third of the nucleon mass. Thus we have
$\FomNN\approx0$ and $\FrhoNN\approx4\GrhoNN$.

With the same prescription, we can obtain similar relations for
heavy charmed baryons which are collected in the appendix.
Substituting the coupling constants at the quark level with those
from Eq.~(\ref{coupNN}), we have
\begin{eqnarray}
\GpiXX=0,\;\; \GpiSS=\frac{4}{5}\GpiNN\frac{m_{\S_c}}{m_N},\;\;
\GpiXXp=\frac{2}{5}\GpiNN\frac{m_{\Xi_c^\p}}{m_N},
\end{eqnarray}
\begin{eqnarray}
\GetaLL=0,\;\; \GetaXX=0,\;\;
\GetaSS=\frac{4}{3}\GetaNN\frac{m_{\S_c}}{m_N},\NO\\ \GetaXXp=-\frac{2}{3}\GetaNN\frac{m_{\Xi_c^\p}}{m_N},\;\;
\GetaOO=-\frac{8}{3}\GetaNN\frac{m_{\O_c}}{m_N},
\end{eqnarray}
\begin{eqnarray}
\GsigLL=\frac{2}{3}\GsigNN,\;\; \GsigXX=\frac{2}{3}\GsigNN,\;\;
\GsigSS=\frac{2}{3}\GsigNN,\NO\\ \GsigXXp=\frac{2}{3}\GsigNN,\;\;
\GsigOO=\frac{2}{3}\GsigNN,
\end{eqnarray}
\begin{eqnarray}
\GomLL=\frac{2}{3}\GomNN,\;\;
\FomLL=-\frac{2}{3}\GomNN,\NO\\ \GomXX=\frac{1}{3}\GomNN,\;\;
\FomXX=-\frac{1}{3}\GomNN,\NO\\ \GomSS=\frac{2}{3}\GomNN,\;\;
\FomSS=\frac{2}{3}\GomNN\left(2\frac{m_{\S_c}}{m_N}-1\right),\NO\\ \GomXXp=\frac{1}{3}\GomNN,\;\;
\FomXXp=\frac{1}{3}\GomNN\left(2\frac{m_{\Xi_c^\p}}{m_N}-1\right),
\end{eqnarray}
\begin{eqnarray}
\GrhoXX=\GrhoNN,\;\;
\FrhoXX=-\frac{1}{5}\left(\GrhoNN+\FrhoNN\right),\NO\\ \GrhoSS=2\GrhoNN,\;\;
\FrhoSS=\frac{2}{5}\left(\GrhoNN+\FrhoNN\right)\left(2\frac{m_{\S_c}}{m_N}-1\right),\NO\\ \GrhoXXp=\GrhoNN,\;\;
\FrhoXXp=\frac{1}{5}\left(\GrhoNN+\FrhoNN\right)\left(2\frac{m_{\Xi_c^\p}}{m_N}-1\right),
\end{eqnarray}
\begin{eqnarray}
\GphiXX=\sqrt{2}\GrhoNN,\;\;
\FphiXX=-\frac{\sqrt{2}}{5}\left(\GrhoNN+\FrhoNN\right),\NO\\ \GphiXXp=\sqrt{2}\GrhoNN,\;\;
\FphiXXp=\frac{\sqrt{2}}{5}\left(\GrhoNN+\FrhoNN\right)\left(2\frac{m_{\Xi_c^\p}}{m_N}-1\right),\NO\\ \GphiOO=2\sqrt{2}\GrhoNN,\;\;
\FphiOO=\frac{2\sqrt{2}}{5}(\GrhoNN+\FrhoNN)\left(2\frac{m_{\O_c}}{m_N}-1\right),\label{coup-phi}
\end{eqnarray}
where we have used $m_N\approx3m_q$.  The couplings of $\phi$ and
heavy charmed baryons can not be derived directly from the results
for nucleons. So in the right hand side of Eq.~(\ref{coup-phi}),
we use the couplings of $\rho$ and nucleons.

The above formula relate the unknown coupling constants for heavy
charmed baryons to $\GpiNN$, $\GetaNN$, etc. which can be
determined by fitting to experimental data. We choose the values
$\GpiNN=13.07$, $\GetaNN=2.242$, $\GsigNN=8.46$, $\GomNN=15.85$,
$\FomNN/\GomNN=0$, $\GrhoNN=3.25$ and $\FrhoNN/\GrhoNN=6.1$ from
Refs.~\cite{Mach87,Mach01,Cao2010} as inputs. In Table~\ref{Tab:coup}, we list the
numerical results of the coupling constants of the heavy charmed
baryons and light mesons. One notices that the vector meson
couplings for $\Xi_c\Xi_c$ and $\L_c\L_c$ have opposite signs.
They almost cancel out and do not contribute to the tensor terms
for spin-triplets. Thus in the following numerical analysis, we
omit the tensor forces of the spin-triplets in the $\Xi_c\Xi_c$
and $\L_c\L_c$ systems.

\begin{center}
\begin{table}[htb]
\begin{tabular}{c|cc|cc|cc|cc|cc}
\hline\hline
&\multicolumn{2}{c|}{$\L_c\L_c$}&\multicolumn{2}{c|}{$\Xi_c\Xi_c$}&\multicolumn{2}{c|}
	{$\S_c\S_c$}&\multicolumn{2}{c|}{$\Xi_c^{'}\Xi_c^{'}$}&\multicolumn{2}{c}{$\O_c\O_c$}\\ \hline
	$\a$~&~$g_{\a \L_c\L_c}$ & $f_{\a\L_c\L_c}$ &~$g_{\a\Xi_c\Xi_c}$ &
	$f_{\a\Xi_c\Xi_c}$ &~$g_{\a\S_c\S_c}$ & $f_{\a\S_c\S_c}$
	&~$g_{\a\Xi_c^{'}\Xi_c{'}}$ & $f_{\a\Xi_c^{'}\Xi_c^{'}}$
	&~$g_{\a\O_c\O_c}$ & $f_{\a\O_c\O_c}$\\ \hline
	$\pi$&~~&~~&~0~&~~&~$27.36$~&~~&~$14.36$~&~~&~~&~\\
	$\eta$&~$0$~&~~&~$0$~&~~&~$7.82$~&~~&~$-4.10$~&~~&~$-17.19$~&~\\
	$\s$&~$5.64$~&~~&~$5.64$~&~~&~$5.64$~&~~&~$5.64$~&~~&~$5.64$~&~\\
	$\o$&~$10.57$~&~$-10.57$~&~$5.28$~&~$-5.28$~&~$10.57$~&~$44.67$~&~$5.28$~&~$23.72$~&~~&~\\
	$\rho$&~~&~~&~$3.25$~&$-4.62$~&~$6.50$~&~$39.01$~&~$3.25$~&~$20.72$~&~~&~~\\
	$\phi$&~~&~~&~$4.60$~&~$-6.53$~&~~&~~&~$4.60$~&~$29.30$~&~$9.19$~&~$61.94$\\ \hline\hline
\end{tabular}
\caption{Numerical results of the coupling constants.  The coupling
  constants with the $\phi$ exchange are deduced from $g_{\rho NN}$.\label{Tab:coup}}
\end{table}
\end{center}

\section{Numerical Results\label{numer}}

With the effective potentials and the coupling constants derived
in the previous sections, one can calculate the binding energies
and root-mean-square (RMS) radii for every possible molecular
state numerically.  Here we adopt the program FESSDE which is a
FORTRAN routine to solve problems of multi-channel coupled
ordinary differential equations~\cite{fessde}. Besides the
coupling constants in Table~\ref{Tab:coup}, we also need heavy
charmed baryon masses listed in Table~\ref{Tab:mass} as inputs.
The typical value of this cutoff parameter for the deuteron is
$1.2\sim1.5\gev$~\cite{Mach87}. In our case, the cutoff parameter $\L$
is taken in the region $0.80\sim2.00\gev$. Such a region is broad
and reasonable enough to give us a clear picture of the
possibility of the heavy baryon molecules.
\begin{widetext}
\begin{center}
\begin{table}[htb]
\begin{tabular}{lclc|lclc}
\hline\hline baryon & mass($\mbox{MeV}$) & baryon &
mass($\mbox{MeV}$)&~meson & mass($\mbox{MeV}$) & meson &
mass($\mbox{MeV}$) \\ \hline $\L_c^+$ & $2286.5$ & $\S_c$ & $2455$
&~$\pi^\pm$ & $139.6$ & $\rho$ & $775.5$ \\ $\Xi_c^+$ & $2467.8$ &
$\Xi_c^{\p+}$ & $2575.6$ &~$\pi^0$ & $135.0$ & $\o$ &
$782.7$\\ $\Xi_c^0$ & $2470.9$ & $\Xi_c^{\p0}$ & $2577.9$ &~$\eta$ &
$547.9$ &$\phi$ & $1019.5$\\ $\O_c^0$ & $2695.2$ & & &~$\s$ & $600$ & &
\\ \hline\hline
\end{tabular}
\caption{Masses of heavy baryons and light
  mesons \cite{pdg2010}.  We use
  $m_{\Xi_c}=2469.3\mev$, $m_{\Xi_c^\p}=2576.7\mev$ and $m_\pi=138.1\mev$ as numerical analysis inputs.\label{Tab:mass}}
\end{table}
\end{center}
\end{widetext}

\subsection{$\Lambda_c\Lambda_c$ and $\Xi_c\Xi_c$ systems}

\begin{figure}[htb]
	  \hfill
	  \begin{minipage}[b]{0.33\textwidth}
	  \centering
	  \includegraphics[width=0.95\textwidth]{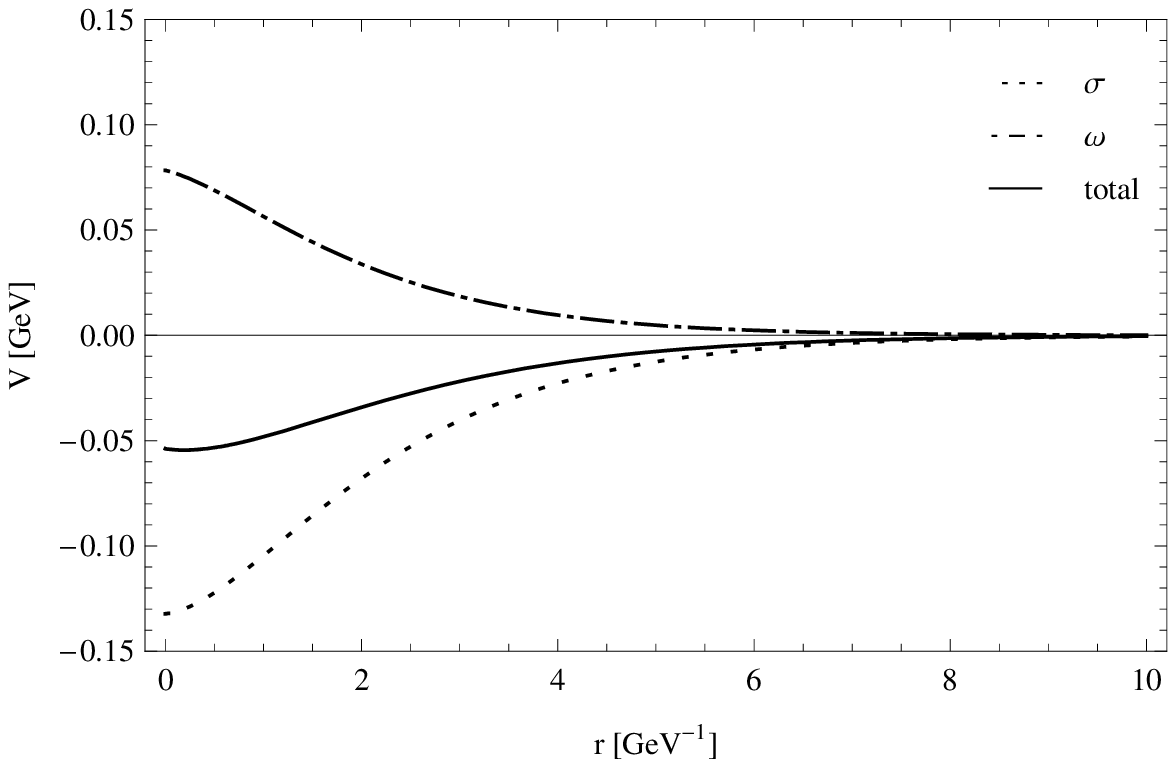}\\ (a) $V_{\L_c\L_c}^{[0,1]}$ with $\L=0.9\gev$.\\
	  \includegraphics[width=0.95\textwidth]{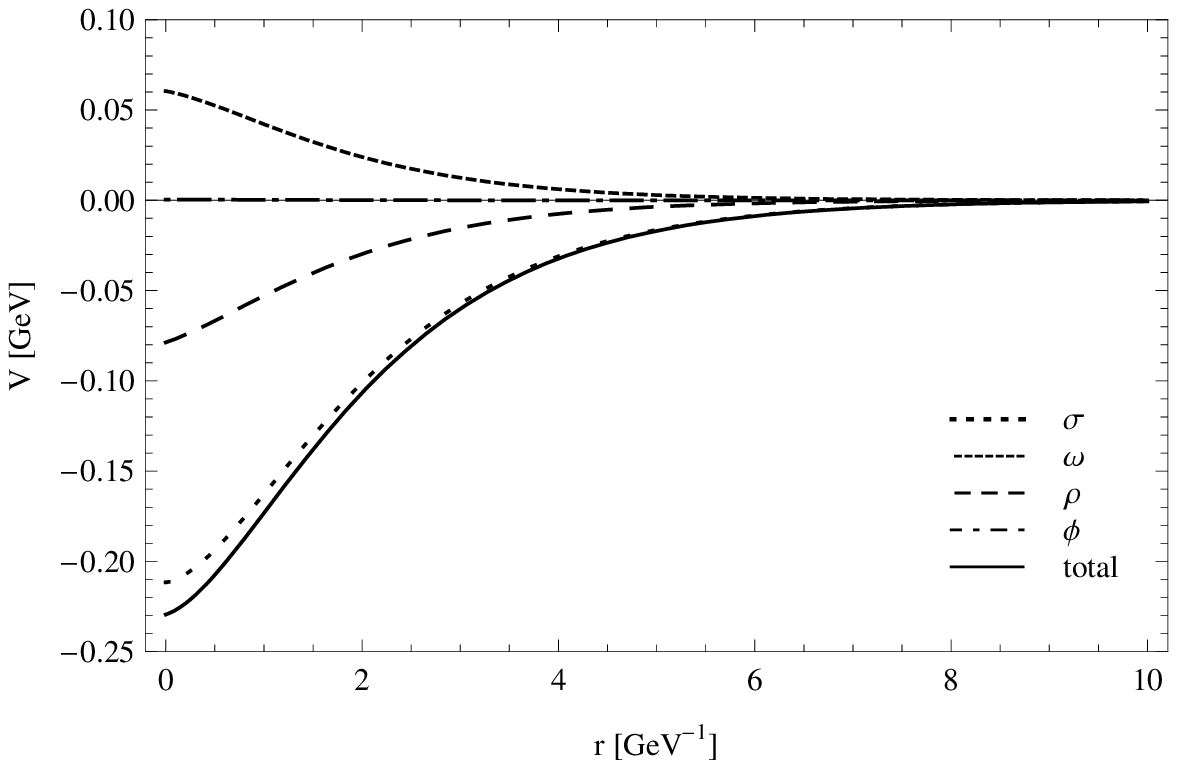}\\ (d) $V_{\Xi_c\Xi_c}^{[0,3]}$ with $\L=1.0\gev$.\\
	\end{minipage}%
	\hfill
	  \begin{minipage}[b]{0.33\textwidth}
	  \centering
	  \includegraphics[width=0.95\textwidth]{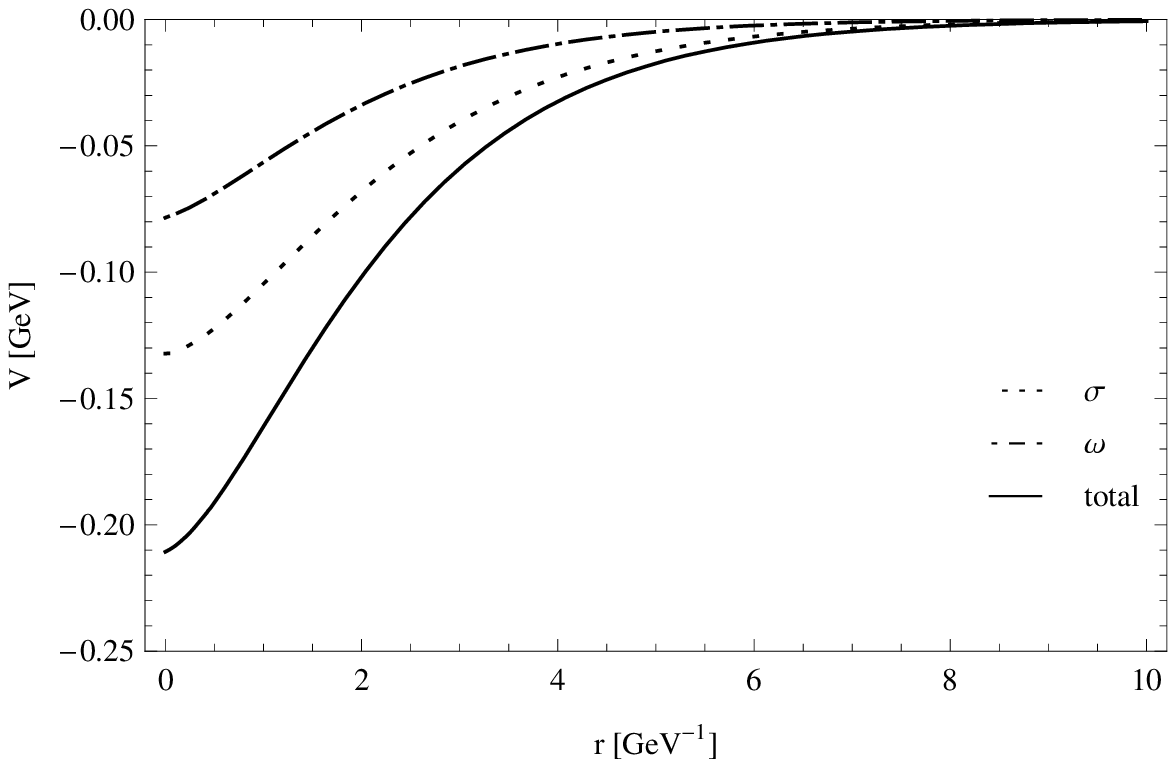}\\ (b) $V_{\L_c\bar{\L}_c}^{[0,1]}$ with $\L=0.9\gev$.\\
	  \includegraphics[width=0.95\textwidth]{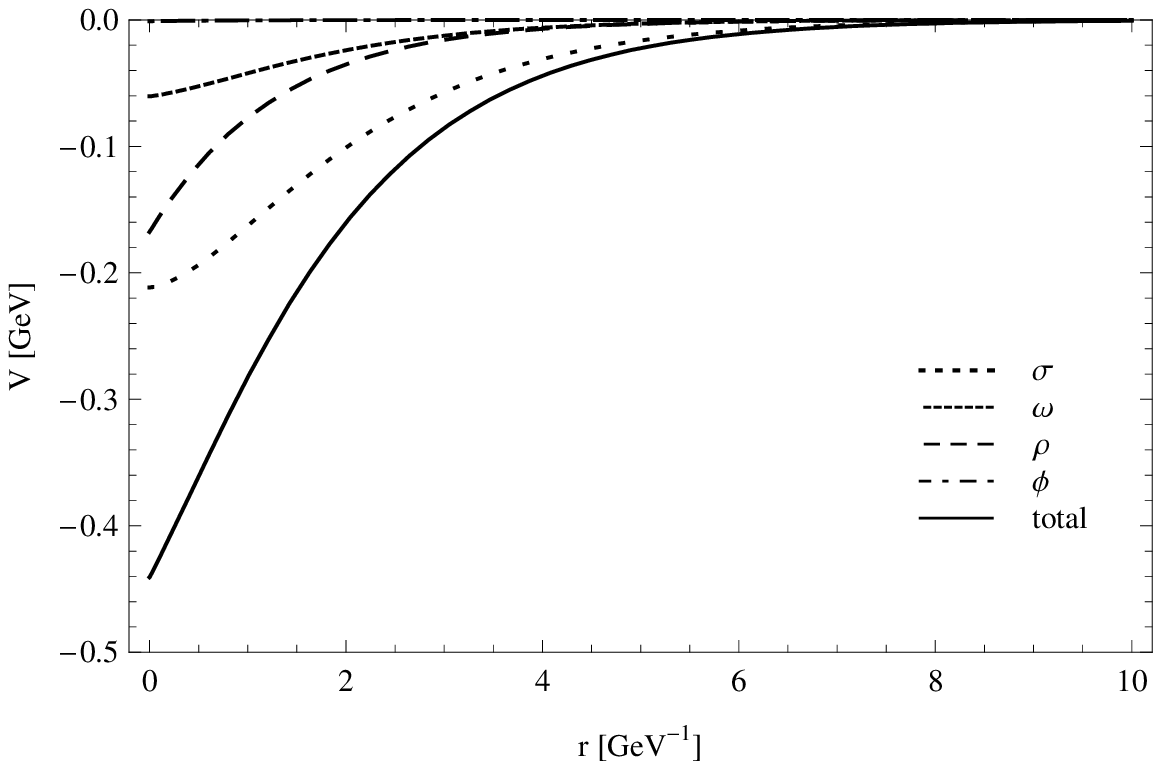}\\ (e) $V_{\Xi_c\bar{\Xi}_c}^{[0,1]}$ with $\L=1.0\gev$.\\
			  \end{minipage}%
	  \begin{minipage}[b]{0.33\textwidth}
	  \centering
	  \includegraphics[width=0.95\textwidth]{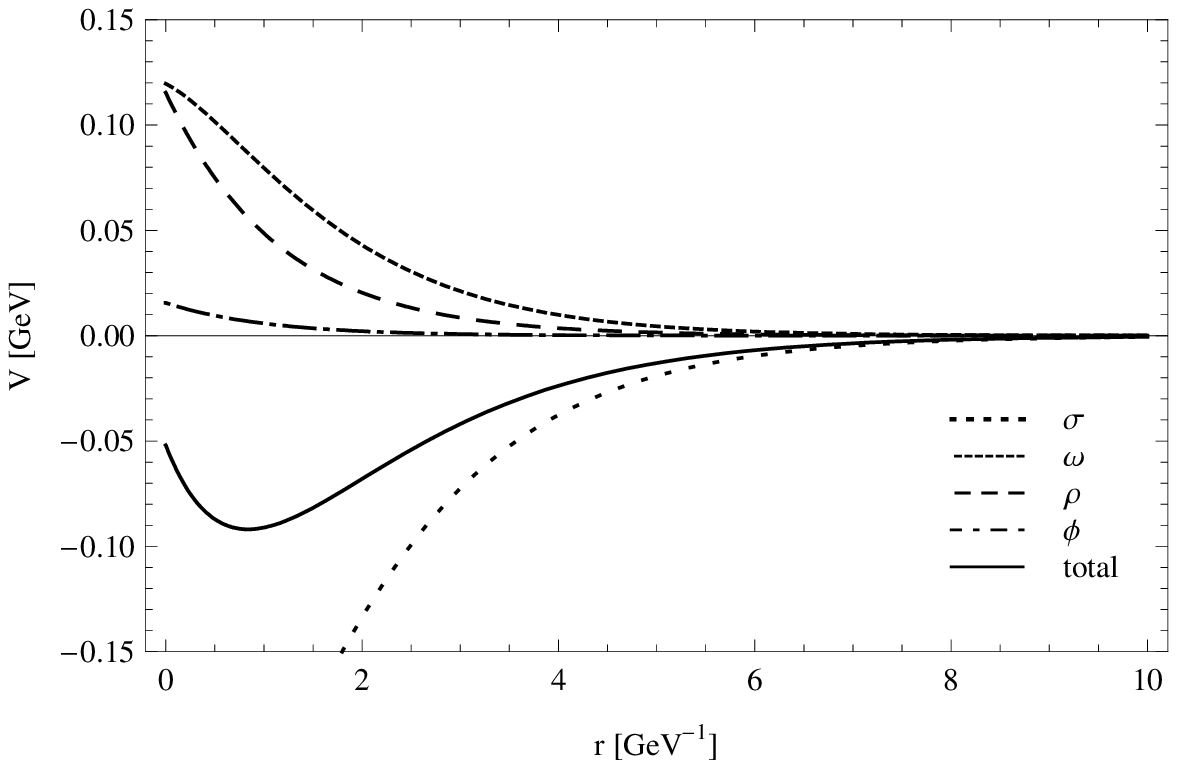}\\ (c) $V_{\Xi_c\Xi_c}^{[1,1]}$ with $\L=1.1\gev$.\\
	  \includegraphics[width=0.95\textwidth]{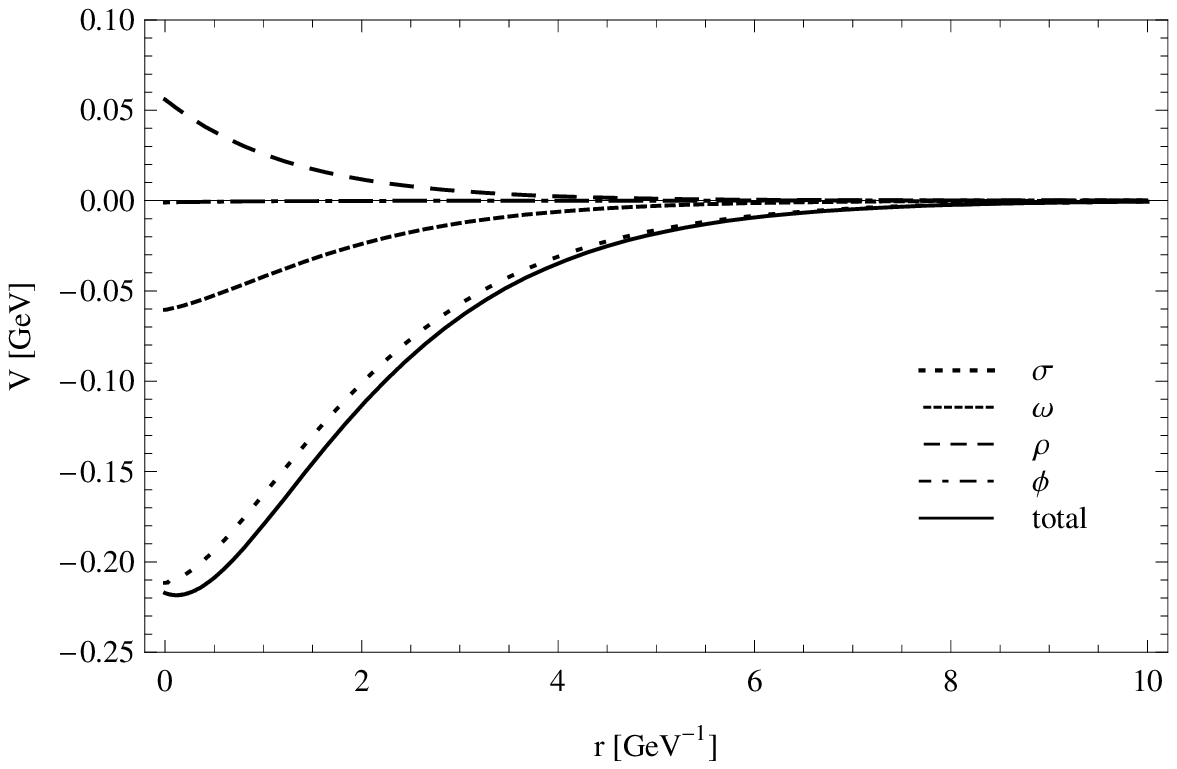}\\ (f) $V_{\Xi_c\bar{\Xi}_c}^{[1,3]}$ with $\L=1.0\gev$.\\
	  \end{minipage}
			  \hspace*{\fill}
  \caption{The potentials of $\Psi_{\L_c\L_c}$, $\Psi_{\L_c\bar{\L}_c}$, $\Psi_{\Xi_c\Xi_c}$
  and $\Psi_{\Xi_c\bar{\Xi}_c}$. The spin-triplets have no $S-D$ mixing because of the cancellations
  of the coupling constants.}
  \label{Fig:Pot3} 
\end{figure}

The total effective potential of $\L_c\L_c$ arises from the $\s$
and $\o$ exchanges. We plot it with $\L=0.9\gev$ in
Fig.~\ref{Fig:Pot3} (a), from which we can see that the $\o$
exchange is repulsive while the $\s$ exchange is attractive.
Because of the cancellation, the total potential is too shallow to
bind two $\L_c$s. In fact, we fail to find any bound solutions of
$\Psi_{\L_c\L_c}^{[0,1]}$ even if one takes the deepest potential
with $\L=0.9\gev$. In other words, the loosely bound $\L_c\L_c$
molecular state does not exist, which is the heavy analogue of the
famous H
dibaryon~\cite{Aerts:1983hy,Iwasaki:1987db,Stotzer:1997vr,Ahn:1996hw}
to some extent.

For the $\L_c\bar{\L}_c$ system as shown in Fig.~\ref{Fig:Pot3}
(b), both $\s$ and $\o$ exchanges are attractive. They enhance
each other and lead to a very strong total interaction. From our
results listed in Table~\ref{Tab:3B3Num}, the binding energies of
the $\L_c\bar{\L}_c$ system could be rather large. For example,
when we increase the cutoff to $\L=1.10\gev$, the corresponding
binding energy is $142.19\mev$. The binding energies and RMS
radii of this system are very sensitive to the cutoff, which
seems to be a general feature of the systems composed of one
hadron and anti-hadron.

\begin{center}
\begin{table}[htb]
\begin{tabular}{|cccc|cccc|}
\hline\hline
~  &~$\Lambda$ (GeV)~~&~~E (MeV)~~&~~$r_{rms}$(fm)~~&~~&~$\Lambda$ (GeV)~~&~E (MeV)~~&~~$r_{rms}$(fm)~\\ \hline
~							  &   &   &   &									 & 0.89&
2.80  & 2.15 \\
$\Psi_{\Lambda_c\Lambda_c}^{[0,1]}$&	  &$-$&   &$\Psi_{\Lambda_c\bar{\Lambda}_c}^{[0,1(3)]}$& 0.90&
4.61 & 1.76 \\
						 ~   &  &	  &   &									 & 1.00&
49.72& 0.74 \\
						 ~   &  &   &   &									& 1.10&
142.19 & 0.52 \\
\hline
~					  & 0.95 & 2.53   & 2.17 &					 & 1.01& 0.14 & 5.58 \\
$\Psi_{\Xi_c\Xi_c}^{[0,3]}$& 1.00 & 7.41   & 1.41 &$\Psi_{\Xi_c\Xi_c}^{[1,1]}$& 1.05& 0.29 & 4.48 \\
						 ~ & 1.10 & 20.92  & 0.96 &			   & 1.10& 0.35 & 4.62 \\
						 ~ & 1.20 & 36.59  & 0.78 &			   & 1.20& 0.18 & 5.40 \\
\hline
								   &0.87 &  1.48&  2.72&								   & 0.90 &
1.24 & 2.92 \\
$\Psi_{\Xi_c\bar{\Xi}_c}^{[0,1(3)]}$&0.90 &  4.12&  1.78&$\Psi_{\Xi_c\bar{\Xi}_c}^{[1,1(3)]}$& 1.00 &
10.33& 1.25\\
								   &1.00 & 28.94&  0.86&								   & 1.10 &
31.80& 0.83 \\
								   &1.10 & 82.86&  0.60&								   & 1.20 &
66.19& 0.64 \\
\hline\hline
\end{tabular}
\caption{Numerical results of the systems $\Lambda_c\Lambda_c$,
$\Lambda_c\bar{\Lambda}_c$, $\Xi_c\Xi_c$ and $\Xi_c\bar{\Xi}_c$,
where ``$-$'' means no bound state solutions. After neglecting the
tensor force terms, the results of spin-triplets are the same as
those of spin-singlets.}\label{Tab:3B3Num}
\end{table}
\end{center}

$\Xi_c^0$ and $\Xi_c^+$ contain the $s$ quark and their isospin is
$I=1/2$. Besides the $\s$ and $\o$ meson exchanges, the $\phi$ and
$\rho$ exchanges also contribute to the potentials for the
$\Xi_c\Xi_c(\bar{\Xi}_c)$ systems. Figs.~\ref{Fig:Pot3} (c) and
(d) illustrate the total potentials and the contributions from the
light meson exchanges for $\Psi_{\Xi_c\Xi_c}^{[1,1]}$ and
$\Psi_{\Xi_c\Xi_c}^{[0,3]}$.  For $\Psi_{\Xi_c\Xi_c}^{[1,1]}$, the
attraction arises from the $\s$ exchange. Because of the repulsion
provided by the $\phi$, $\rho$ and $\o$ exchange in short range,
the total potential has a shallow well at $r\approx0.2\fm$.
However, the $\phi$ exchange almost does not contribute to the
potential of $\Psi_{\Xi_c\Xi_c}^{[0,3]}$ and the $\rho$ exchange
is attractive which cancels the repulsion of the $\s$ exchange. The
total potential is about two times deeper than the total potential
of $\Psi_{\Xi_c\Xi_c}^{[1,1]}$.

In Table~\ref{Tab:3B3Num}, one notices that the binding energy is
only hundreds of $\kev$ for $\Psi_{\Xi_c\Xi_c}^{[1,1]}$ when
the cutoff varies from $1.01\gev$ to $1.20\gev$. Moreover the RMS
radius of this bound state is very large. So the state
$\Psi_{\Xi_c\Xi_c}^{[1,1]}$ is very loosely bound if it really
exists. The $\Psi_{\Xi_c\Xi_c}^{[0,3]}$ bound state may also
exist. Its binding energy and RMS radius are $2.53\sim36.59\mev$
and $2.17\sim0.78\fm$ respectively with $\L=0.95\sim1.20\gev$.

As for the $\Xi_c\bar{\Xi}_c$ systems, the potentials are very
deep. The contribution from the $\phi$ exchange is negligible too,
as shown in Fig.~\ref{Fig:Pot3} (e) and (f). We find four bound
state solutions for these systems:
$\Psi_{\Xi_c\bar{\Xi}_c}^{[0,1]}$,
$\Psi_{\Xi_c\bar{\Xi}_c}^{[0,3]}$,
$\Psi_{\Xi_c\bar{\Xi}_c}^{[1,1]}$ and
$\Psi_{\Xi_c\bar{\Xi}_c}^{[1,3]}$. Among them, the numerical
results of $\Psi_{\Xi_c\bar{\Xi}_c}^{[0,3]}$ and
$\Psi_{\Xi\bar{\Xi}_c}^{[1,1]}$ are almost the same as those of
$\Psi_{\Xi\bar{\Xi}_c}^{[0,1]}$ and
$\Psi_{\Xi_c\bar{\Xi}_c}^{[1,3]}$ respectively. The binding
energies and the RMS radii of these states are shown in
Table~\ref{Tab:3B3Num}. We can see that the binding energy of
$\Psi_{\Xi_c\bar{\Xi}_c}^{[0,1]}$ varies from $1.48\mev$ to
$82.86\mev$ whereas the RMS radius reduces from $2.72\fm$ to
$0.60\fm$ when the cutoff is below $1.10\gev$. The situation of
$\Psi_{\Xi_c\bar{\Xi}_c}^{[1,3]}$ is similar to that of
$\Psi_{\Xi_c\bar{\Xi}_c}^{[0,1]}$ qualitatively. They may exist.
But the binding energies appear a little large and the RMS radii
too small when one takes $\L$ above $1.10\gev$.

\subsection{$\Sigma_c\Sigma_c$, $\Xi^\prime_c\Xi^\prime_c$ and  $\Omega_c\Omega_c$ systems}

\begin{figure}[htb]
	  \hfill
	  \begin{minipage}[t]{0.33\textwidth}
	  \centering
	  \includegraphics[width=0.95\textwidth]{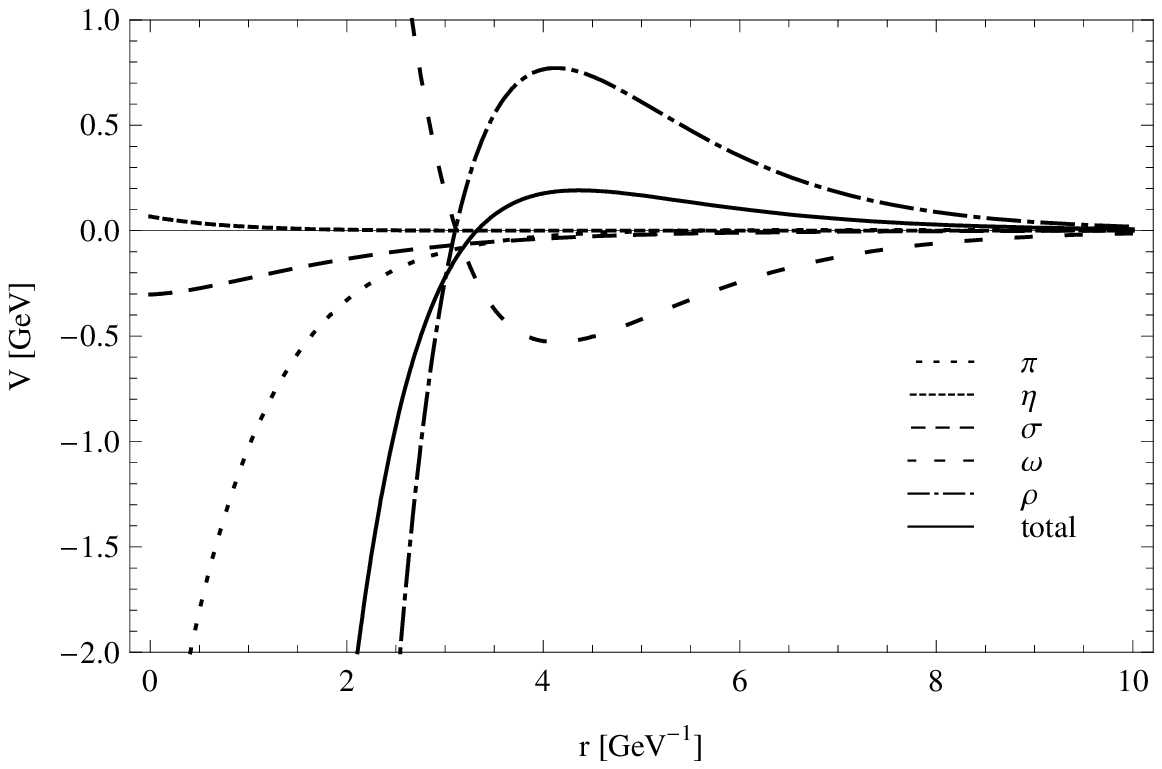}\\ (a) $V_{\S_c\S_c}^{[0,1]}$ with $\L=1.10\gev$.\\
	  \includegraphics[width=0.95\textwidth]{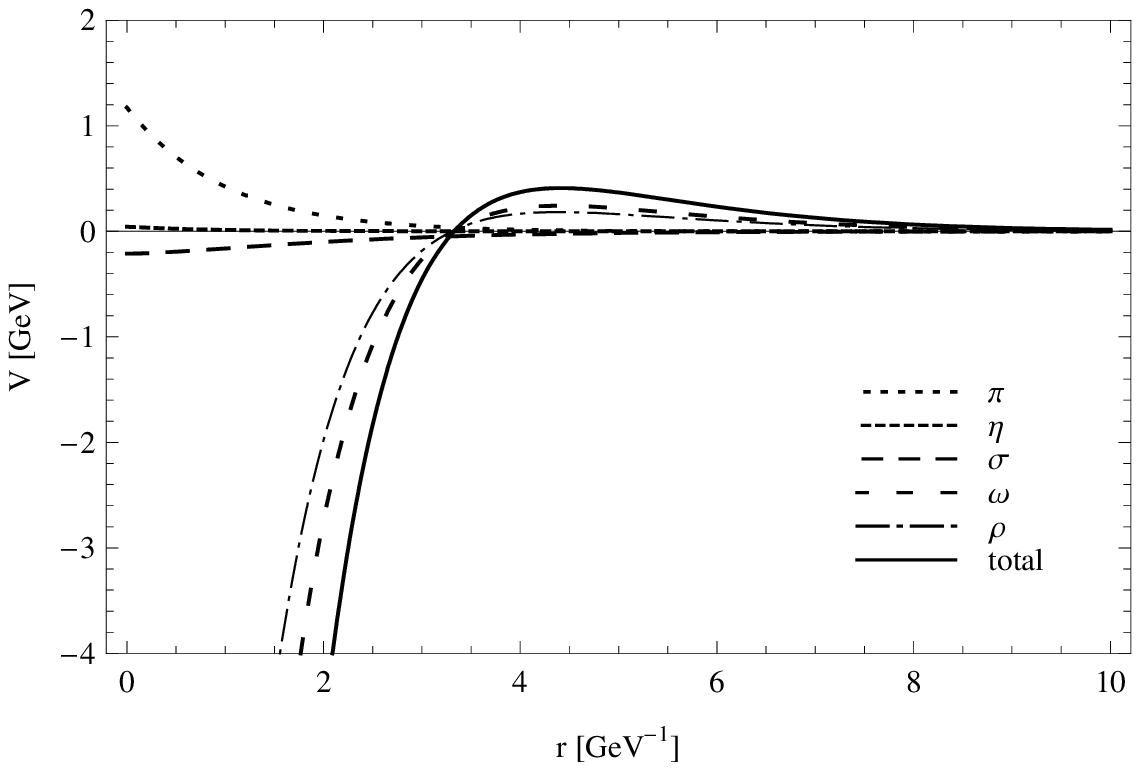}\\ (d) $V_{\S_c\bar{\S}_c}^{[1,1]}$ with $\L=1.0\gev$.\\
	  \includegraphics[width=0.95\textwidth]{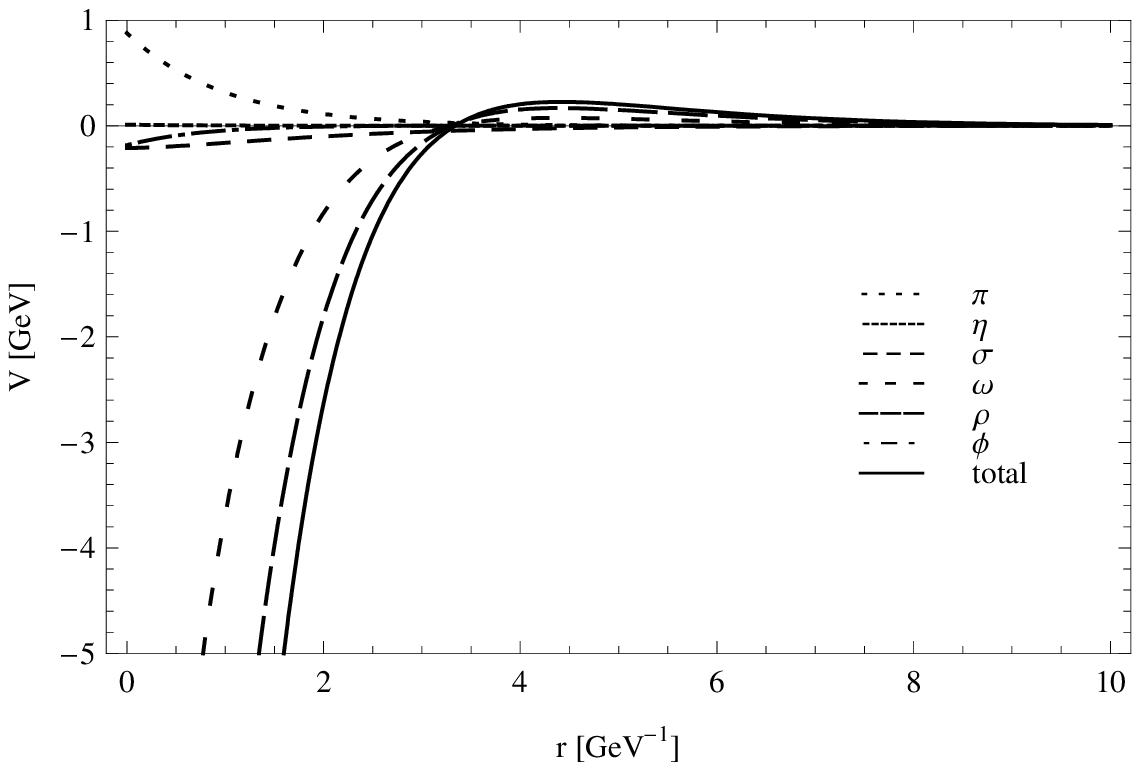}\\ (g) $V_{\Xi_c^\p\bar{\Xi}_c^\p}^{[0,1]}$ with $\L=1.0\gev$.\\
	  \includegraphics[width=0.95\textwidth]{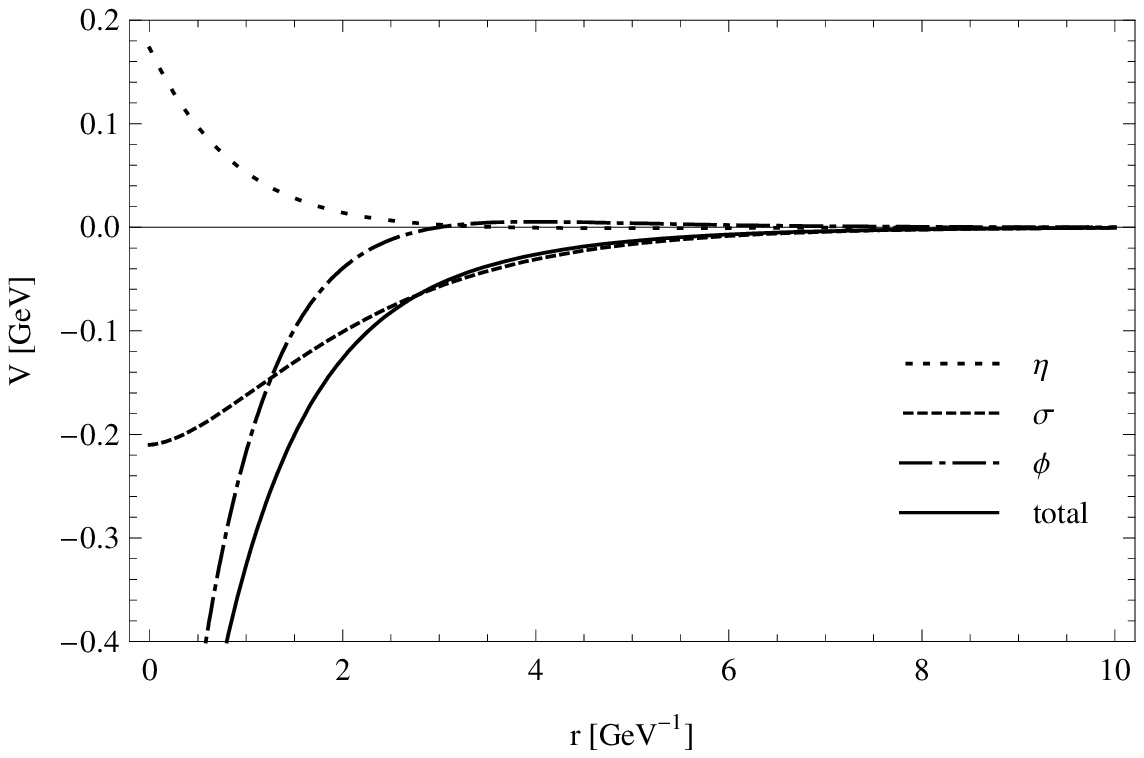}\\ (j) $V_{\O_c\bar{\O}_c}^{[0,1]}$ with $\L=1.0\gev$.\\
	\end{minipage}%
	\hfill
	  \begin{minipage}[t]{0.33\textwidth}
	  \centering
	  \includegraphics[width=0.95\textwidth]{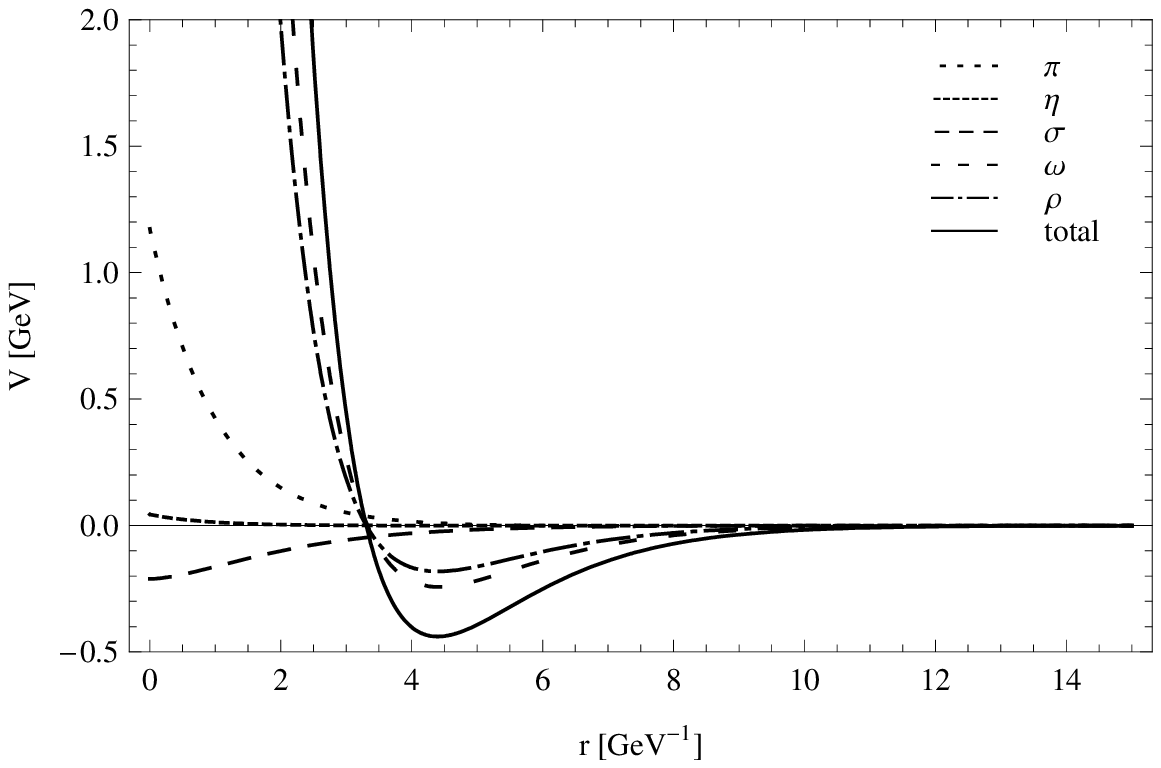}\\ (b) $V_{\S_c\S_c}^{[2,1]}$ with $\L=1.0\gev$.\\
	  \includegraphics[width=0.95\textwidth]{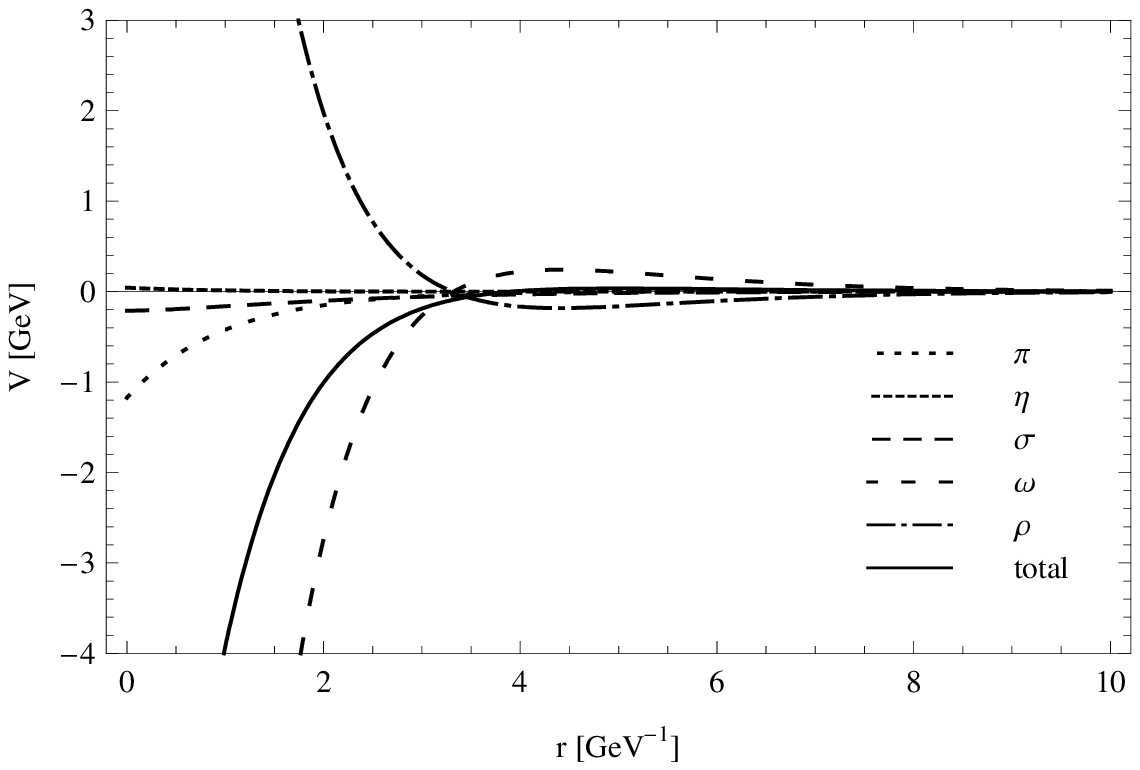}\\ (e) $V_{\S_c\bar{\S}_c}^{[2,1]}$ with $\L=1.0\gev$.\\
	  \includegraphics[width=0.95\textwidth]{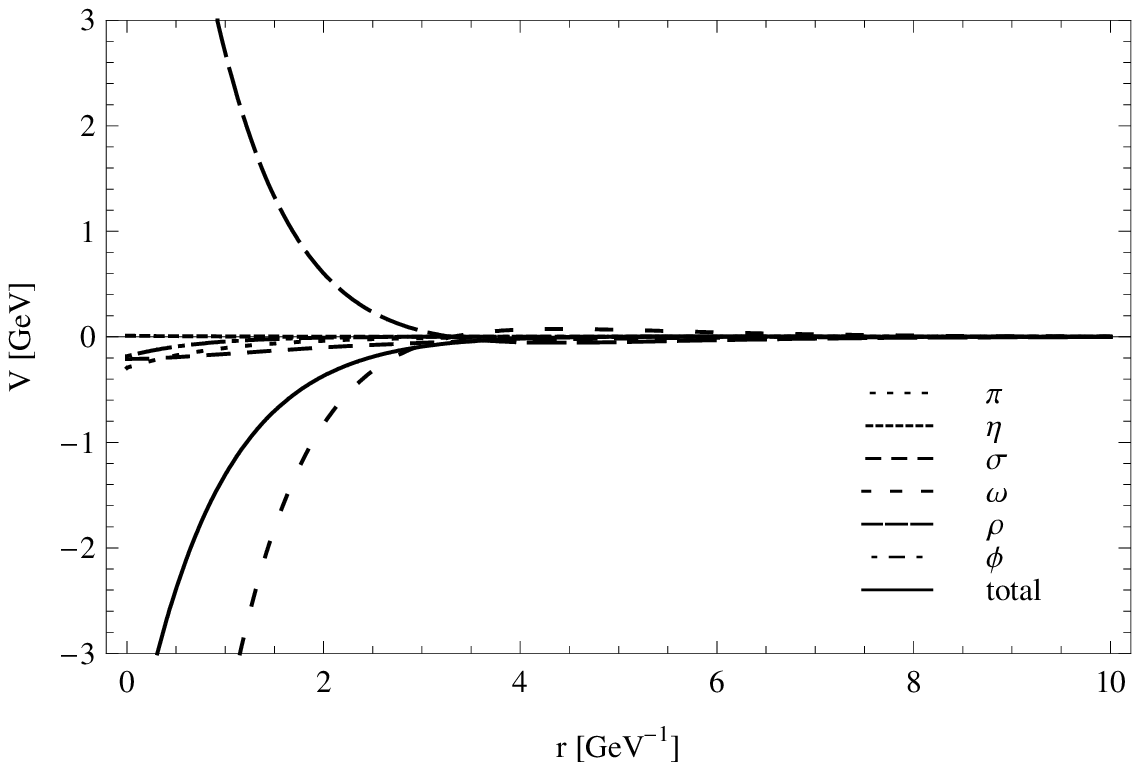}\\ (h) $V_{\Xi_c^\p\bar{\Xi}_c^\p}^{[1,1]}$ with $\L=1.0\gev$.\\
			  \end{minipage}%
	  \begin{minipage}[t]{0.33\textwidth}
	  \centering
	  \includegraphics[width=0.95\textwidth]{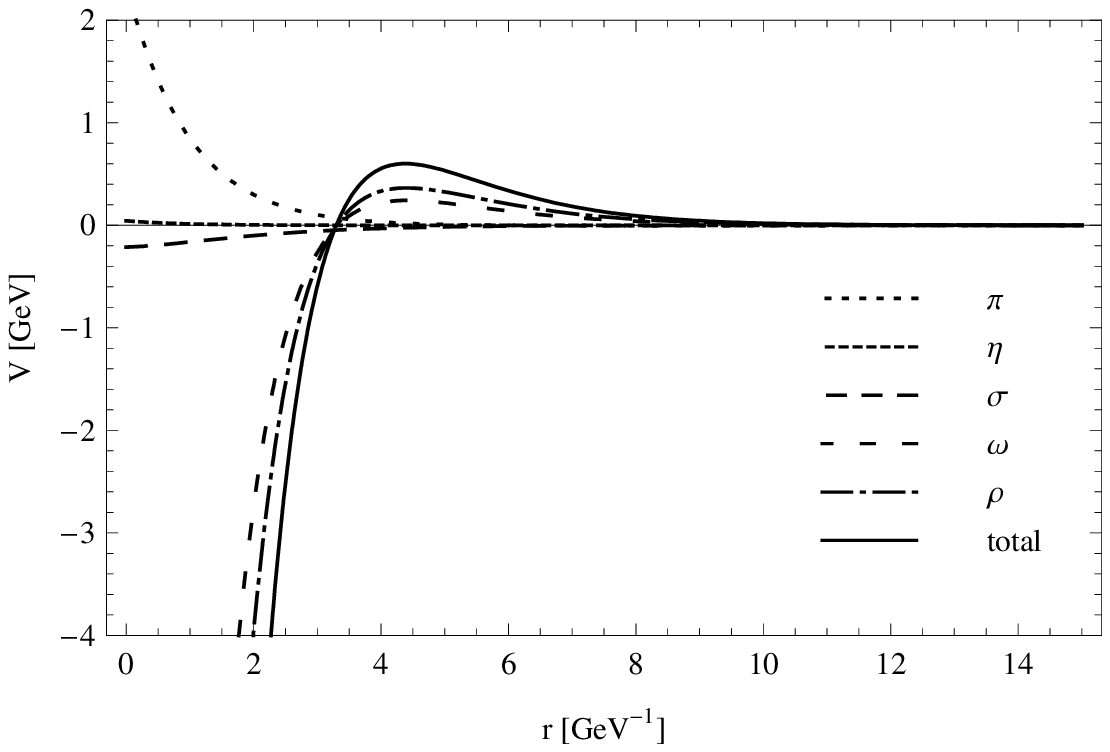}\\ (c) $V_{\S_c\bar{\S}_c}^{[0,1]}$ with $\L=1.0\gev$.\\
	  \includegraphics[width=0.95\textwidth]{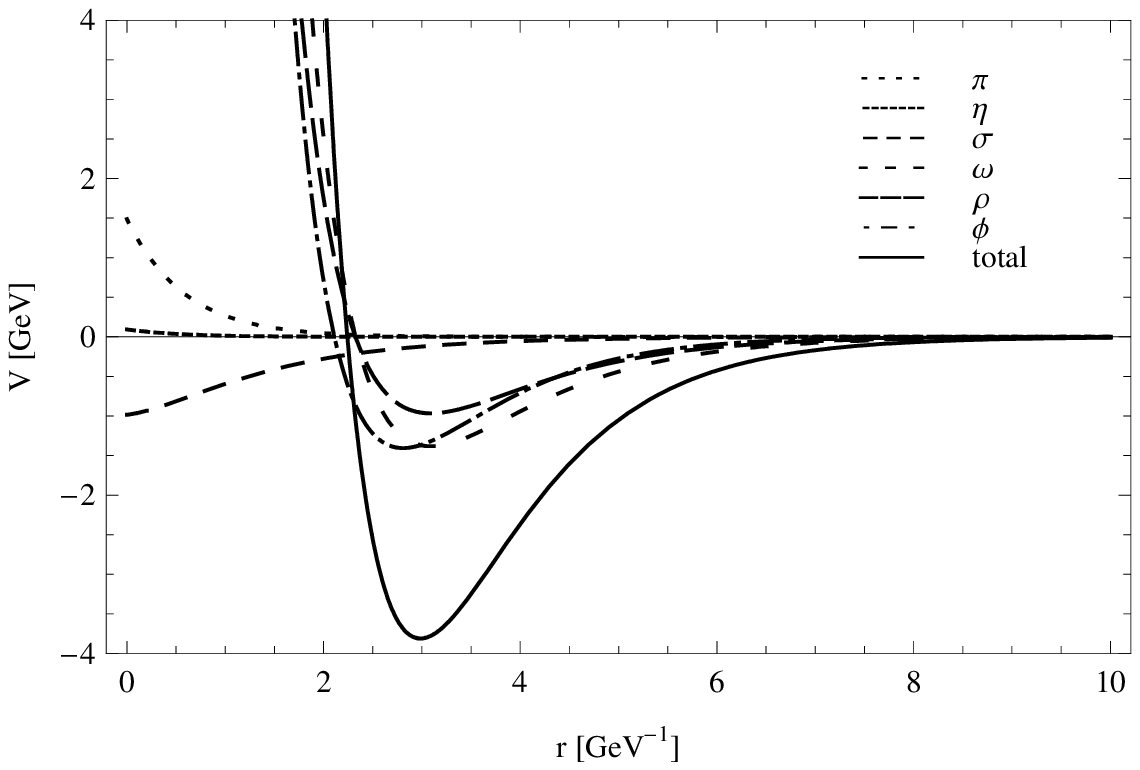}\\ (f) $V_{\Xi_c^\p\Xi_c^\p}^{[1,1]}$ with $\L=1.70\gev$.\\
	  \includegraphics[width=0.95\textwidth]{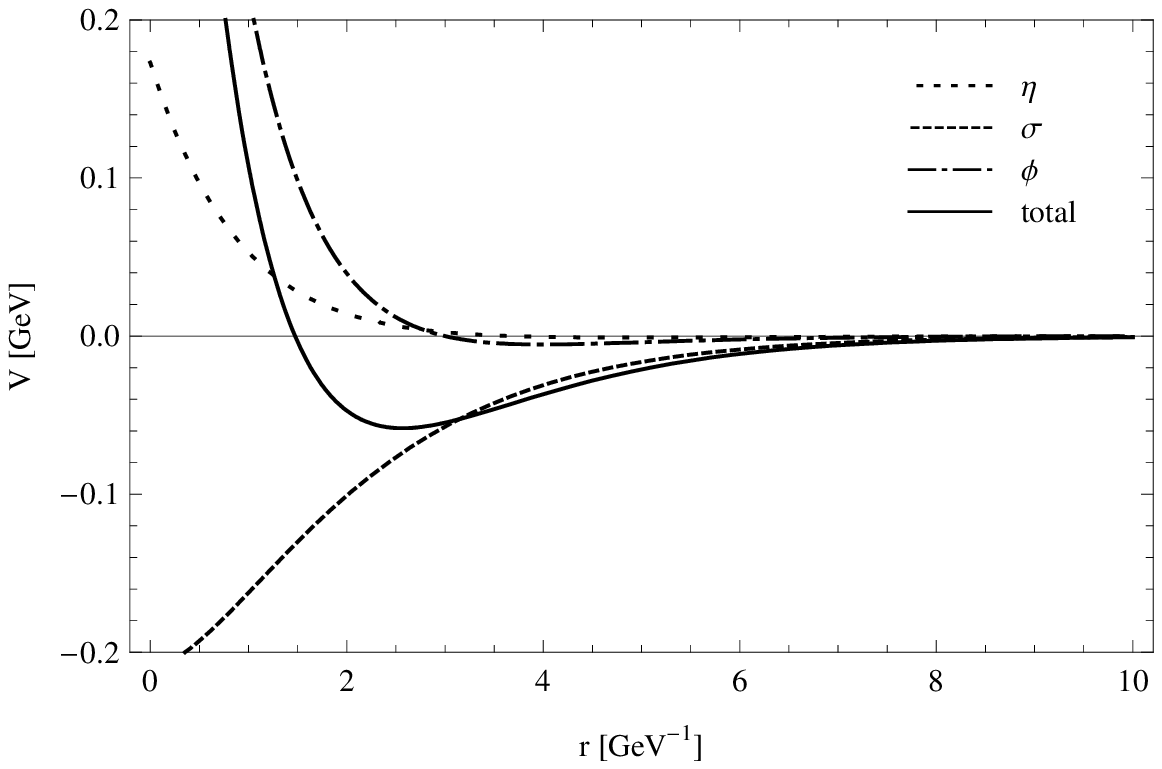}\\ (i) $V_{\O_c\O_c}^{[0,1]}$ with $\L=1.0\gev$.\\
	  \end{minipage}
			  \hspace*{\fill}
  \caption{Potentials of $\Psi_{\S_c\S_c}$, $\Psi_{\S_c\bar{\S}_c}$, $\Psi_{\Xi_c^\p\Xi_c^\p}$,
   $\Psi_{\Xi_c^\p\bar{\Xi}_c^\p}$, $\Psi_{\O_c\O_c}$ and $\Psi_{\O_c\bar{\O}_c}$. }
  \label{Fig:Pot6} 
\end{figure}

For the $\S_c\S_c$ system, all the $\pi$, $\eta$, $\s$, $\o$ and
$\rho$ exchanges contribute to the total potential. We give the
variation of the potentials with $r$ in Figs.~\ref{Fig:Pot6} (a)
and (b). For $\Psi_{\S_c\S_c}^{[0,1]}$, the potential of the $\o$
exchange and $\rho$ exchange almost cancel out, and the $\eta$
exchange gives very small contribution. So the total potential of
this state mainly comes from the $\pi$ and $\s$ exchanges which
account for the long and medium range attraction respectively.
There may exist a bound state $\Psi_{\S_c\S_c}^{[0,1]}$, see
Table~\ref{Tab:SSNum}.

But for the other spin-singlet, $\Psi_{\S_c\S_c}^{[2,1]}$, the
$\s$ exchange provides only as small as $0.2\gev$ attraction
while the $\o$ and $\rho$ exchanges give strong repulsions in
short range $r<0.6\fm$. We have not found any bound solutions for
$\Psi_{\S_c\S_c}^{[2,1]}$ as shown in Table~\ref{Tab:SSNum}. For
the spin-triplet state $\Psi_{\S_c\S_c}^{[1,3]}$, there exist bound
state solutions with binding energies between $0.11\mev$ and
$31.35\mev$ when the cutoff lies between $1.05\gev$ and
$1.80\gev$. This state is the mixture of $^3S_1$ and $^3D_1$ due
to the tensor force in the potential. From Table~\ref{Tab:SSNum},
one can see the $S$ wave percentage is more than $90\%$.

\begin{center}
\begin{table}[htb]
\begin{tabular}{|c|ccc|c|cccll|}
\hline\hline
~&$\Lambda$ (GeV)&E (MeV)&$r_{rms}$ (fm)&~&$\Lambda$ (GeV)&~E (MeV)~&~$r_{rms}$ (fm)~&~
$P_S$~:&$P_D$ (\%) \\ \hline
				~			&1.07 & 1.80 & 2.32 &  ~  &   &   &			&   &   \\
				~			&1.08 & 3.10 & 1.88 &  ~  &   &   &			&   &   \\
$\Psi_{\Sigma_c\Sigma_c}^{[0,1]}$&1.10 & 6.55 & 1.44 &  $\Psi_{\Sigma_c\Sigma_c}^{[0,3]}$&&$\times$&&&\\
				~			&1.20 &42.95 & 0.78 &  ~  &   &   &		  &   & \\
				~			&1.25 &75.75 & 0.65 &  ~  &   &   &		  &   & \\
\hline
				~			& &   &   & ~ & 1.05  & 0.11 & 5.94  &98.11 & 1.89  \\
$\Psi_{\Sigma_c\Sigma_c}^{[1,1]}$&   &   &   &$\Psi_{\Sigma_c\Sigma_c}^{[1,3]}$& 1.47  & 2.03 & 2.48  &94.21 & 5.79  \\
				~			& &$\times$&& ~ & 1.50  & 2.52 & 2.27  &93.79 & 6.21  \\
				~			& &   &   & ~ & 1.80  & 31.35& 0.76  & 91.41& 8.59  \\
\hline
				~			& &   &   & ~ &   & &  & &   \\
$\Psi_{\Sigma_c\Sigma_c}^{[2,1]}$&   &$-$&  &$\Psi_{\Sigma_c\Sigma_c}^{[2,3]}$& &$\times$&  &   &	\\
				~			& &   &   & ~ &   & &   &   & \\
\hline\hline
\end{tabular}
\caption{Numerical results of the $\Sigma_c\Sigma_c$ system, where
the symbol ``$\times$'' means this state is forbidden and ``$-$''
means no solutions.}\label{Tab:SSNum}
\end{table}
\end{center}

\begin{center}
\begin{table}[htb]
\begin{tabular}{|c|cccll|cccll|}
\hline\hline
~&\multicolumn{5}{c|}{One Boson Exchange}&\multicolumn{5}{c|}{One Pion Exchange}\\ \hline
~&~$\Lambda$(GeV)~&~E (MeV)~&~$r_{rms}$(fm)~&~$P_S$~~:&$P_D$(\%)~&~$\Lambda$(GeV)~&~E(MeV)~&~$r_{rms}$(fm)~&~
$P_S$~~:&$P_D$(\%) \\
\hline
				~				  &0.97 & 0.86 & 3.76 &  ~ & &   &   &	  &   & \\
$\Psi_{\Sigma_c\bar{\Sigma}_c}^{[0,1]}$&0.98 & 3.03 & 2.21 &	&   &   &   &	   &   & \\
									   &1.00 & 18.43& 1.01 &  ~ &   &   &   &$-$&   &  \\
				~				  &1.05 &175.56& 0.41 &  ~ & &   &   &	  &   & \\
\hline
				~				  &0.93 &1.04  & 3.50&81.20&18.80&0.80&17.54&   1.20 &82.93  & 17.07 \\
$\Psi_{\Sigma_c\bar{\Sigma}_c}^{[0,3]}$&0.94 &2.55  & 2.57&75.27&24.73&0.85&26.33&   1.04   &81.66  & 18.34 \\
									   &1.00 &28.16 & 1.29&58.07&41.93&0.90&37.48&   0.92   &80.57  & 19.42\\
				~				  &1.05 &78.48 & 0.99&50.56&49.44&1.05&87.94&   0.68 &78.03  & 21.97  \\
\hline
				~				  &0.93 &0.75  & 3.77&   ~ &  &  &   &	 &   &\\
$\Psi_{\Sigma_c\bar{\Sigma}_c}^{[1,1]}$&0.94 &2.54  & 2.27&  &   &  &   &$-$&   &\\
									   &0.98 &32.28 & 0.80&  &   &  &   &	   &  &\\
				~				  &1.00 &66.97 & 0.60& ~   &  &  &   &	  &  &\\
\hline

				~				  &0.80 & 3.71 & 1.91 &94.73&5.27&0.97&1.04 &   3.14  &93.68  & 6.32\\
$\Psi_{\Sigma_c\bar{\Sigma}_c}^{[1,3]}$&0.81 & 5.18 & 1.69 &94.38&5.62&1.02&2.51 &   2.18   &91.58  & 8.42 \\
									   &0.90 & 40.35& 0.86 &90.12&9.88&1.10&6.44 &   1.51   &89.04  & 10.96 \\
				~				  &1.00 &143.46& 0.62 &76.86&23.14&1.30&27.27&  0.88  &84.89  & 15.11\\
\hline
				~				  &0.80 &24.87 & 0.85 &  &   &0.75&2.49&   1.98 &  &   \\
$\Psi_{\Sigma_c\bar{\Sigma}_c}^{[2,1]}$&0.85 &49.30 & 0.67 &	 &   &0.80&5.95&   1.38 &	& \\
									   &0.90 &90.04 & 0.55 &	 &   &0.90&18.30&  0.88 &	&  \\
				~				  &0.95 &149.66&0.46  &  &   &1.10&72.23&  0.51 &  &   \\
\hline
				~				  &0.90 & 1.44 & 2.93 &96.92& 3.08& & &	   &	&  \\
$\Psi_{\Sigma_c\bar{\Sigma}_c}^{[2,3]}$&1.00 & 14.99& 1.21 &95.43& 4.57&	&   &$-$&   &  \\
									   &1.10 & 41.81& 0.86 &95.11&4.89 &	&   &	   &  &  \\
				~				  &1.20 & 77.28& 0.71 &94.72& 5.28& & &		&  &  \\
\hline\hline
\end{tabular}
\caption{Numerical results of the $\S_c$-$\bar{\S}_c$ system.
Results from the OBE and OPE alone are
compared.}\label{Tab:SBSNum}
\end{table}
\end{center}

There exist bound state solutions for all six states of the
$\S_c\bar{\S}_c$ system. The potentials of the three spin-singlets
are plotted in Figs. ~\ref{Fig:Pot6} (c)-(e). The attraction that
binds the baryonium mainly comes from the $\rho$ and $\o$
exchanges. These contributions are of relatively short range at
region $r<0.6\fm$. One may wonder whether the annihilation of the
heavy baryon and anti-baryon might play a role here. Thus the
numerical results for $\S_c\bar{\S}_c$ with strong short-range
attractions should be taken with caution. This feature differs
from the dibaryon systems greatly.

In Table~\ref{Tab:SBSNum}, for comparison, we also present the
numerical results with the $\pi$ exchange only. It's very
interesting to investigate whether the long-range
one-pion-exchange potential (OPE) alone is strong enough to bind the
baryonia and form loosely bound molecular states. There do
not exist bound states solutions for $\Psi_{\S_c\bar{\S}_c}^{[0,1]}$ and
$\Psi_{\S_c\bar{\S}_c}^{[1,1]}$ since the $\pi$ exchange is
repulsive. In contrast, the attractions from the $\pi$ exchange are
strong enough to form baryonium bound states for
$\Psi_{\S_c\bar{\S}_c}^{[0,3]}$, $\Psi_{\S_c\bar{\S}_c}^{[1,3]}$
and $\Psi_{\S_c\bar{\S}_c}^{[2,1]}$. We notice that the $S-D$
mixing effect for the spin-triplets mentioned above is stronger
than that for the $\S_c\S_c$ system.

\begin{center}
\begin{table}[htb]
\begin{tabular}{|c|ccc|ccccll|}
\hline\hline
~&$\Lambda$ (GeV)~&~E (MeV)~&~$r_{rms}$ (fm)~&~&$\Lambda$ (GeV)~&~E (MeV)~&~$r_{rms}$ (fm)~&~~
$P_S$~:&$P_D$ (\%) \\ \hline
				 ~		 &   &		&  &						   &0.95
& 1.22 & 3.03 & 97.88&2.12  \\
				 ~		 &   &		&  &							&0.98
& 2.44 & 2.29 & 97.45&2.55  \\
$\Psi_{\Xi_c^{'}\Xi_c^{'}}^{[0,1]}$&	 &$\times$&   &$\Psi_{\Xi_c^{'}\Xi_c^{'}}^{[0,3]}$&1.00
& 3.41 & 2.01 & 97.26&2.74  \\
				 ~		 &   &		&  &							&1.20
& 15.43& 1.16 & 96.74&3.26  \\
				 ~		 &   &		&  &							&1.30
& 21.50& 1.03 & 96.83&3.17  \\
\hline
				 ~		 &1.50 & 0.18   & 5.52 &							&
&   &   &   &   \\
				 ~		 &1.65 & 1.24   & 3.08 &							&
&   &   &   &   \\
$\Psi_{\Xi_c^{'}\Xi_c^{'}}^{[1,1]}$&1.70 & 1.83 & 2.64 &$\Psi_{\Xi_c^{'}\Xi_c^{'}}^{[1,3]}$&
&   & $\times$ &   & \\
				 ~		 &1.80 & 3.42   & 2.08 &							&
&   &   &   &   \\
				 ~		 &1.90 & 5.58   & 1.74 &							&
&   &   &   &   \\
\hline\hline
\end{tabular}
\caption{Numerical results of the $\Xi_c^{'}\Xi_c^{'}$
system.}\label{Tab:XPXPNum}
\end{table}
\end{center}

\begin{center}
\begin{table}[htb]
\begin{tabular}{|ccccll|cccll|}
\hline\hline
~&\multicolumn{5}{c|}{One Boson Exchanges}&\multicolumn{5}{c|}{One Pion Exchanges}\\
\hline
~&$\Lambda$ (GeV)~&~E (MeV)~&~$r_{rms}$ (fm)~&~$P_s$~~:&$P_D$(\%)&$\Lambda$ (GeV)~&~E (MeV)~&~$r_{rms}$ (fm)~&~~
$P_S$~:&$P_D$ (\%) \\ \hline
~									&0.96 & 0.40 & 4.57 &  &
&   &  &	 &   & \\
										 &0.99 & 3.22 & 2.00 &  &
&   &  &$-$&	 &   \\
$\Psi_{\Xi_c^{'}\bar{\Xi}_c^{'}}^{[0,1]}$&1.00 & 5.13 & 1.65 &  &
&   &  &		   &   &   \\
										 &1.10 &83.53 & 0.58 &  &
&   &  &	 &   & \\
\hline
										 & 0.80 & 3.82 & 1.86 &  96.33  &  3.67
& 1.15  & 0.77 & 3.42 & 94.89 & 5.11  \\
										 & 0.90 & 19.40 &1.04 & 94.34   &  5.66
&1.20 & 1.89 & 2.35 & 93.01 & 6.99  \\
$\Psi_{\Xi_c^{'}\bar{\Xi}_c^{'}}^{[0,3]}$& 1.00 & 59.74 & 0.74 & 90.03  &  9.97
&1.40  & 12.69 & 1.10 & 88.10 & 11.90  \\
										 & 1.05 & 90.87 & 0.66 &86.20  &   13.80
&1.50  & 22.91& 0.88 & 86.44 & 13.56  \\
\hline
										 & 0.80 & 14.13 & 1.01 &		&
&   &   &   &  &	   \\
$\Psi_{\Xi_c^{'}\bar{\Xi}_c^{'}}^{[1,1]}$& 0.90 & 13.58 & 1.07 &		&
&   &   &$-$&   &   \\
										 & 1.00 & 34.00 & 0.77 &		&
&   &   &		&   &   \\
										 & 1.10 & 83.78& 0.56 &  &
&   &   &		&   &   \\
\hline
										 &0.90  & 0.56 & 3.99 & 99.76  & 0.24
&   &   &		&   &   \\
$\Psi_{\Xi_c^{'}\bar{\Xi}_c^{'}}^{[1,3]}$&1.00  & 7.53 & 1.41 & 99.59  &  0.41
&   &   &$-$&   &   \\
										 &1.10  & 22.97& 0.94 & 99.58  &  0.42
&   &   &		&   &   \\
										 & 1.20 & 43.80& 0.76 & 99.58  &  0.42
&   &   &		&   &   \\
\hline\hline
\end{tabular}
\caption{Comparison of the numerical results of the system
$\Xi_c^{'}\bar{\Xi}_c^{'}$ in the OBE model and OPE
model.}\label{Tab:XPBXPNum}
\end{table}
\end{center}

The $\Xi_c^\p\Xi_c^\p(\bar{\Xi}_c^\p)$ systems are similar to
$\Xi_c\Xi_c(\bar{\Xi}_c)$ and the results are listed in
Figs.~\ref{Fig:Pot6} (f)-(h) and
Tables~\ref{Tab:XPXPNum}-\ref{Tab:XPBXPNum}. Among the six bound
states, $\Psi_{\Xi_c^\p\Xi_c^\p}^{[1,1]}$ is the most interesting
one. As shown in Fig.~\ref{Fig:Pot6} (f), the $\eta$ exchange does
not contribute to the total potential. The $\pi$ exchange is
repulsive. So the dominant contributions are from the $\s$, $\o$,
$\rho$ and $\phi$ exchanges, which lead to a deep well
around $r=0.6\fm$ and a loosely bound state. When we
increase the cutoff from $1.50\gev$ to $1.90\gev$, the binding
energy of $\Psi_{\Xi_c^\p {\Xi}_c^\p}^{[1,1]}$ varies from
$0.18\mev$ to $5.58\mev$, and the RMS radius varies from $5.52$ fm
to $1.74\fm$. This implies the existence of this loosely bound
state. If we consider the $\pi$ exchange alone, only the
$\Psi_{\Xi_c^\p\bar{\Xi}_c^\p}^{[0,3]}$ state is bound. The
percentage of the $^3S_1$ component is more than $86\%$ when
$1.15\gev < \L < 1.50\gev$ as shown in Table~\ref{Tab:XPBXPNum}.

\begin{center}
\begin{table}[htb]
\begin{tabular}{|cccc|ccccll|}
\hline\hline
~  & $\Lambda$ (GeV)~&~E (MeV)~&~$r_{rms}$ (fm)~&~&$\Lambda$ (GeV)~&~E (MeV)~&~$r_{rms}$ (fm)~&~~
$P_S$~:&$P_D$ (\%) \\ \hline
								  & 0.96 & 1.07 & 3.04 &								 &
&	&  &   &  \\
								  & 0.98 & 2.67 & 2.08 &								 &
&	&  &   &  \\
$\Psi_{\Omega_c\Omega_c}^{[0,1]}$ & 1.00 & 4.51 & 1.69 &$\Psi_{\Omega_c\Omega_c}^{[0,3]}$&
&$\times$&  &   & \\
								  & 1.20 & 5.92 & 1.59 &								 &
&	&  &   &   \\
								  & 1.70 & 19.88& 1.15 &								 &
&	&  &   &   \\
\hline
									   & 0.90 & 13.12& 1.06 &								   &0.80
&  6.92 &  1.53 & 99.64&  0.06\\
									   & 0.97 & 4.34 & 1.70 &								   &0.88
&  3.05 &  1.98 & 99.96 & 0.04 \\
$\Psi_{\Omega_c\bar{\Omega}_c}^{[0,1]}$&1.00  & 5.01 & 1.62 &$\Psi_{\Omega_c\bar{\Omega}_c}^{[0,3]}$&1.00
&  9.77 &  1.23 & 99.90 & 0.10 \\
									   &1.10 &  20.96& 0.94 &								   &1.10
& 26.22 &  0.86 & 99.79 &  0.21\\
									   &1.20 & 108.50& 0.48 &								   &1.20
&47.23 &   0.72 & 99.53 &  0.47 \\
\hline\hline
\end{tabular}
\caption{Numerical results of the $\Omega_c\Omega_c$ and
$\Omega_c\bar{\Omega}_c$ systems.}\label{Tab:OONum}
\end{table}
\end{center}

The $\O_c\O_c(\bar{\O}_c)$ case is quite simple. Only the $\eta$,
$\s$ and $\phi$ exchanges contribute to the total potentials. The
shape of the potential of $\Psi_{\O_c\O_c}^{[0,1]}$ is similar to
that of $\Psi_{\Xi_c^\p\Xi_c^\p}^{[1,1]}$. The binding energy of
this state is very small. For the spin-triplet $\O_c\bar{\O}_c$
system, its $S$ wave percentage is more than $99\%$. In other
words, the $S-D$ mixing effect is tiny for this system.

We give a brief comparison of our results with
those of Refs.~\cite{Froemel:2004ea,JuliaDiaz:2004rf} in
Table~\ref{Tab:Riska}. In Ref.~\cite{Froemel:2004ea}, Fr\"oemel et
al. deduced the potentials of nucleon-hyperon and hyperon-hyperon
by scaling the potentials of nucleon-nucleon. With the
nucleon-nucleon potentials from different models, they discussed
possible molecular states such as $\Xi_{cc}N$, $\Xi_c\Xi_{cc}$,
$\S_c\S_c$ etc.. The second column of Table~\ref{Tab:Riska} shows
the binding energies corresponding different models while the last
column is the relevant results of this work. One can see the
results of Ref.~\cite{Froemel:2004ea} depend on models while our
results are sensitive to the cutoff $\L$.

\begin{center}
\begin{table}[htb]
\begin{tabular}{c|cccccccccc|c}
\hline \hline
 models	& Nijm93   &   NijmI	 &   NijmII	& AV18		 &  AV$8^{\prime}$
		 & AV$6^\prime$& AV$4^\prime$   & AV$X^{\prime}$& AV$2^\prime$ & AV$1^\prime$
& This work  \\
\hline
$[\Xi_c^{'}\Xi_c^{'}]_{I=0}$ &  -  & *  &	$71.0$  & $457.0$	 &	-
& $0.7$   &   $24.5$	 &  $9.5$	   &	$12.8$  &  -		& $1.22 \sim 21.50$\\
$[\Sigma_c\Sigma_c]_{I=2}$   &   $66.6$   &  - &	  -	 &   $41.1$	& -
&   -   &	-	 &	   -		&	 -	 &   $0.7$	 &   -   \\
$[\Sigma_c\Sigma_c]_{I=1}$   &	 -		&	 *	 &	$53.7$ &   -
&	   -			&	 -		&	 $7.3$  &  $2.8$	 & $8.3$
&	  $0.7$		   &	 $0.11 \sim 31.35$ \\
$[\S_c\S_c]_{I=0}$   &	  *	  &	   *	   &	 $285.8$   &	 *   &
$16.1$				&  $10.8$	  &	$87.4$	&	 $53.3$   &	  $58.5$
& $0.7$			  & $1.80 \sim 75.75$  \\
\hline\hline
\end{tabular}
\caption{The comparison of the binding energies of $\Xi^\p\Xi^\p$
and $\S_c\S_c$ systems in this work and those in Ref.%
~\cite{Froemel:2004ea}.  The unit is$\mev$. ``-''  means there is
no bound state and ``*'' represents exiting unrealistic deeply
bindings ($1\sim10\gev$). }\label{Tab:Riska}
\end{table}
\end{center}

\section{Conclusions\label{summ}}

The one boson exchange model is very successful in the description
of the deuteron, which may be regarded as a loosely bound
molecular system of the neutron and proton. It's very interesting
to extend the same framework to investigate the possible molecular
states composed of a pair of heavy baryons. With heavier mass and
reduced kinetic energy, such a system is non-relativistic. We
expect the OBE framework also works in the study of the heavy
dibaryon system.

On the other hand, one should be cautious when extending the OBE
framework to study the heavy baryonium system. The difficulty lies
in the lack of reliable knowledge of the short-range interaction
due to the heavy baryon and anti-baryon annihilation. However,
there may exist a loosely bound heavy baryonium state when one
turns off the short-range interaction and considers only the
long-range one-pion-exchange potential. Such a case is
particularly interesting. This long-range OPE attraction may lead
to a bump, cusp or some enhancement structure in the heavy baryon
and anti-baryon invariant mass spectrum when they are produced in
the $e^+e-$ annihilation or B decay process etc.

In this work, we have discussed the possible existence of the
$\L_c\L_c(\bar{\L}_c)$, $\Xi_c\Xi_c(\bar{\Xi}_c)$,
$\S_c\S_c(\bar{\S}_c)$, $\Xi^\p_c\Xi^\p_c(\bar{\Xi}_c^\p)$ and
$\O_c\O_c(\bar{\O}_c)$ molecular states. We consider both the long
range contributions from the pseudo-scalar meson exchanges and the
short and medium range contributions from the vector and scalar
meson exchanges.

Within our formalism, the heavy analogue of the
H dibaryon
$\Psi_{\L_c\L_c}^{[0,1]}$ does not exist though its potential is
attractive. However, the $\Psi_{\L_c\bar{\L}_c}^{[0,1]}$ and
$\Psi_{\L_c\bar{\L}_c}^{[0,3]}$ bound states might exist. For the
$\Xi_c\Xi_c$ system, there exists a loosely bound state
$\Psi_{\Xi_c\Xi_c}^{[1,1]}$ with a very small binding energy and a
very large RMS radius around $5\fm$. The spin-triplet state
$\Psi_{\Xi_c\Xi_c}^{[0,3]}$ may also exist. Its binding energy and
RMS radius vary rapidly with increasing cutoff $\L$. The
qualitative properties of $\Psi_{\Xi_c\bar{\Xi}_c}^{[0,1]}$ and
$\Psi_{\Xi_c\bar{\Xi}_c}^{[1,3]}$ are similar to those of
$\Psi_{\L_c\bar{\L}_c}^{[0,1]}$. They could exist but the binding
energies and RMS radii are unfortunately very sensitive to the
values of the cutoff parameter.

For the $\S_c\S_c$, $\S_c\bar{\S}_c$, $\Xi_c^\p\Xi_c^\p$,
$\Xi_c^\p\bar{\Xi}_c^\p$, $\O_c\O_c$ and $\O_c\bar{\O}_c$ systems,
the tensor forces lead to the $S-D$ wave mixing. There probably
exist the $\S_c\S_c$ molecules $\Psi_{\S_c\S_c}^{[0,1]}$ and
$\Psi_{\S_c\S_c}^{[1,3]}$ only. For the $\S_c\bar{\S}_c$ system,
the $\o$ and $\rho$ exchanges are crucial to form the bound states
$\Psi_{\S_c\bar{\S}_c}^{[0,1]}$, $\Psi_{\S_c\bar{\S}_c}^{[1,1]}$
and $\Psi_{\S_c\bar{\S}_c}^{[2,3]}$. If one considers the $\pi$
exchange only for the $\Xi_c^\p\bar{\Xi}_c^\p$ system, there may
exist one bound state $\Psi_{\Xi_c^\p\bar{\Xi}_c^\p}^{[0,3]}$.

The states $\Psi_{\Xi_c\Xi_c}^{[0,3]}$ and
$\Psi_{\Xi_c^\p\Xi_c^\p}^{[0,3]}$ are very interesting. They are
similar to the deuteron.  Especially, $\Psi_{\Xi_c\Xi_c}^{[0,3]}$
and $\Psi_{\Xi_c^\p\Xi_c^\p}^{[0,3]}$ have the same quantum
numbers as deuteron. For $\Psi_{\Xi_c\Xi_c}^{[0,3]}$, the $S-D$
mixing is negligible whereas for deuteron such an effect can make
the percentage of the $D$ wave up to
$4.25\%\sim6.5\%$~\cite{Mach87,Rijken,SprungEtc}. The $D$ wave
percentage of $\Psi_{\Xi_c^\p\Xi_c^\p}^{[0,3]}$ is
$2.12\%\sim3.17\%$.

The other two states $\Psi_{\Xi_c\Xi_c}^{[1,1]}$ and
$\Psi_{\Xi_c^\p\Xi_c^\p}^{[1,1]}$ are very loosely bound $S$ wave
states. Remember that the binding energy of deuteron is about
$2.22\mev$~\cite{HoukAndLeun} with a RMS radius
$r_{rms}\approx1.96\fm$~\cite{BerardAndSimon}. The binding energy
and RMS radius of $\Psi_{\Xi_c^\p\Xi_c^\p}^{[1,1]}$ is quite close
to those of the deuteron. In contrast, the state
$\Psi_{\Xi_c\Xi_c}^{[1,1]}$ is much more loosely bound. Its
binding energy is only a tenth of that of deuteron.

However, the binding mechanisms for the deuteron and the above
four bound states are very different. For the deuteron, the
attraction is from the $\pi$ and vector exchanges. But for these
four states, the $\pi$ exchange contribution is very small. Either
the $\s$ (for $\Xi_c\Xi_c$) or vector meson (for
$\Xi_c^\p\Xi_c^\p$) exchange provides enough attractions to bind
the two heavy baryons.

Although very difficult, it may be possible to produce the charmed
dibaryons at RHIC and LHC. Once produced, the states $\Xi_c\Xi_c$
and $\Xi_c^\p\Xi_c^\p$ are stable since $\Xi_c$ and $\Xi_c^\p$
decays either via weak or electromagnetic interaction with a
lifetime around $10^{-15}s$~\cite{pdg2010}. On the other hand,
$\S_c$ mainly decays into $\L_c^+\pi$. However its width is only
$2.2\mev$ ~\cite{pdg2010}. The relatively long lifetime of $\S_c$
allows the formation of the molecular states
$\Psi_{\S_c\S_c}^{[0,1]}$ and $\Psi_{\S_c\S_c}^{[0,1]}$. These
states may decay into $\S_c\L_c^+\pi$ or $\L_c^+\L_c^+\pi\pi$ if
the binding energies are less than $131\mev$ or $62\mev$
respectively.  Another very interesting decay mode
is $\Xi_{cc}N$ with the decay momentum around one hundred MeV. In
addition, a baryonium can decay into one charmonium and some light
mesons. In most cases, such a decay mode may be kinetically
allowed. These decay patterns are characteristic and useful to the
future experimental search of these baryonium states.

Up to now, many charmonium-like ``XYZ'' states have been observed
experimentally. Some of them are close to the two charmed meson
threshold. Moreover, Belle collaboration observed a near-threshold
enhancement in $e^+e^-\to\L_c\bar{\L}_c$ ISR process with the mass
and width of $m=(4634^{+8}_{-7}(stat.)^{+5}_{-8}(sys.))\mev/c^2$
and $\Gamma_{tot}=(92^{+40}_{-24}(stat.)^{+10}_{-21}(sys.))\mev$
respectively ~\cite{Pakhlova:2008vn}. BaBar collaboration also
studied the correlated leading $\L_c\bar{\L}_c$ production
\cite{Aubert:2010yz}. Our investigation indicates there does exist
strong attraction through the $\s$ and $\omega$ exchange in the
$\L_c\bar{\L}_c$ channel, which mimics the correlated two-pion and
three-pion exchange to some extent.

Recently, ALICE collaboration observed the production of nuclei
and antinuclei in $pp$ collisions at LHC
~\cite{Collaboration:2011yf}. A significant number of light nuclei
and antinuclei such as (anti)deuterons, (anti)tritons,
(anti)Helium3 and possibly (anti)hypertritons with high statistics
of over $350\mathrm{~M}$ events were produced. Hopefully the heavy
dibaryon and heavy baryon and anti-baryon pair may also be
produced at LHC. The heavy baryon and anti-baryon pair may also be
studied at other facilities such as PANDA, J-Parc and Super-B
factories in the future.

\section*{Acknowledgments}

We thank Profs.~Wei-Zhen Deng, Jun He, Gui-Jun Ding and Jean-Marc Richard for useful
discussions. This project is supported by the National Natural
Science Foundation of China under Grants No.~11075004,
No.~11021092, and the Ministry of Science and Technology of China
(No.~2009CB825200).


\begin{thebibliography}{99}

\bibitem{Belle}
S.~K.~Choi, S.~L.~Olsen, {\em et al.} [Belle Collaboration],
Phys.\ Rev.\ Lett.\ {\bf91}, 262001 (2003);
Phys.\ Rev.\ Lett.\ {\bf94}, 182002 (2005);
Phys.\ Rev.\ Lett.\ {\bf100}, 142001 (2008);
X.~L.~Wang, {\em et al.} [Belle Collaboration],
Phys.\ Rev.\ Lett.\ {\bf99}, 142002 (2007);
P.~Pakhlov, {\em et al.} [Belle Collaboration],
Phys. Rev. Lett. {\bf100}, 202001 (2008).

\bibitem{BaBar}
B.~Aubert, {\em et al.} [BaBar Collaboration],
  Phys.\ Rev.\ Lett.\ {\bf101}, 082001 (2008);
  Phys.\ Rev.\ Lett.\ {\bf95}, 142001 (2005);
  Phys.\ Rev.\ Lett.\ {\bf98}, 212001 (2007).

\bibitem{CDF}
D.~E.~Acosta, T.~Affolder, {\em et al.} [CDF Collaboration],
 Phys.\ Rev.\ Lett.\ {\bf93}, 072001 (2004).

\bibitem{D0}
V.~M.~Abazov, {\em et al.} [D0 Collaboration],
Phys.\ Rev.\ Lett.\ {\bf93}, 162002 (2004).


\bibitem{Brambilla:2010cs}
  N.~Brambilla {\it et al.},
  Eur.\ Phys.\ J.\  C {\bf 71}, 1534 (2011)
  arXiv:1010.5827 [hep-ph].

\bibitem{Swanson2006}
E.~S.~Swanson, Phys.\ Rept.\ {\bf 429}, 243 (2006).



\bibitem{Rujula77}
A.~D.~Rujula, H.~Georgi and S.~L.~Glashow,
 Phys.\ Rev.\ Lett.\ {\bf38}, 317 (1977).

\bibitem{Torq}
N.~A.~Tornqvist, Z.\ Phys.\ C {\bf61}, 525 (1994).


\bibitem{Swan04}
E.~S.~Swanson, Phys.\ Lett.\ B {\bf588}, 189 (2004).

\bibitem{Wong04}
C.~Y.~Wong, Phys.\ Rev.\ C\ {\bf69}, 055202 (2004).

\bibitem{Close2004}
F.~E.~Close and P.~R.~Page, Phys.\ Lett.\ B {\bf578}, 119 (2004).

\bibitem{Voloshin2004}
M.~B.~Voloshin, Phys.\ Lett.\ B {\bf579}, 316 (2004).

\bibitem{Thomas2008}
C.~E.~Thomas and F.~E.~Close, Phys.\ Rev.\ D {\bf78}, 034007 (2008).


\bibitem{LiuXLiuYR}
Y.~R.~Liu, X.~Liu, W.~Z.~Deng and S.~L.~Zhu,
Eur.\ Phy.\ J.\ C {\bf56} 63 (2008);
X.~Liu, Y.~R.~Liu, W.~Z.~Deng and S.~L.~Zhu,
Phys.\ Rev.\ D {\bf77} 094015 (2008);
Phys.\ Rev.\ D {\bf77}, 034003 (2008).

\bibitem{ZhugrpDD}
X.~Liu., Z.~G.~Luo, Y.~R.~Liu and S.~L.~zhu,
Eur.\ Phys.\ J.\ C {\bf61}, 411 (2009);
L.~L.~Shen, X.~L.~Chen, {\em et al.},  Eur.\ Phys.\ J.\  C {\bf 70}, 183 (2010);
B.~Hu, X.~L.~Chen, {\em et al.}, Chin.\ Phys.\  C {\bf 35}, 113 (2011);
X.~Liu and S.~L.~Zhu, Phys.\ Rev.\ D {\bf80}, 017502 (2009);
X.~Liu, Z.~G.~Luo and S.~L.~Zhu, arXiv:1011.1045 [hep-ph].


\bibitem{Ping:2000dx}
  J.~Ping, H.~Pang, F.~Wang and T.~Goldman,
  Phys.\ Rev.\  C {\bf 65}, 044003 (2002)
  [arXiv:nucl-th/0012011].

\bibitem{Ding}
G.~ J.~ Ding, Phys.\ Rev.\ D {\bf80}, 034005 (2009); G.~J.~Ding,
J.~F.~Liu and M.~L.~Yan, Phys.\ Rev.\ D {\bf79}, 054005 (2009).

\bibitem{LiuX}
X.~Liu, Eur.\ Phys.\ J.\ C {\bf54}, 471 (2008).

\bibitem{Liu2009}
F. Huang and Z.~Y.~Zhang, Phys.\ Rev.\ C {\bf 72}, 068201 (2005);
Y. R. Liu and Z. Y. Zhang, Phys.\ Rev.\ C {\bf79}, 035206 (2009);
Q. B. Li, P. N. Shen, Z. Y. Zhang and Y. W. Yu, Nucl. Phys. A 683,
487 (2001).

\bibitem{Liu:2011xc}
  Y.~R.~Liu and M.~Oka,
  arXiv:1103.4624 [hep-ph].

\bibitem{qiao}Y. D. Chen and C. F. Qiao, arXiv:1102.3487 [hep-ph]

\bibitem{Mach87}
R.~Machleidt, K.~Holinde and C.~Elster,
Phys.\ Rept.\ {\bf149}, 1 (1987).

\bibitem{Mach01}
R.~Machleidt, Phys.\ Rev.\ C {\bf63}, 024001 (2001).

\bibitem{Rijken}
M.~M.~Nagels, T.~A.~Rijken and J.~D.~de Swart,
 Phys.\ Rev.\  D {\bf12} 744 (1975);
 Phys.\ Rev.\ D {\bf15} 2547 (1977).

\bibitem{Yan}
T.~M.~Yan, {\em et al.} Phys.\ Rev.\ D {\bf46}, 1148 (1992).

\bibitem{Barnes:1999hs}
  T.~Barnes, N.~Black, D.~J.~Dean and E.~S.~Swanson,
  Phys.\ Rev.\  C {\bf 60}, 045202 (1999)
  [arXiv:nucl-th/9902068].

\bibitem{IGrule}
E.~Klempt, F.~Bradamante, {\em et al.} Phys.\ Rept.\ {\bf368} (2002)

\bibitem{Riska2001}
D.~O.~Riska and G.~E.~Brown, Nucl.\ Phys.\ A {\bf679}, 577 (2001).

\bibitem{Cao2010}
X.~Cao, B.~S.~Zou and H.~S.~Xu, Phys.\ Rev.\ C {\bf81}, 065201 (2010).




\bibitem{fessde}
A.~G.~Abrashkevich, D.~G.~Abrashkevich,
M.~S.~Kaschiev and I.~V.~Puzynin, Comput.\ Phys.\ Commun.\ {\bf85} 65-81 (1995).

\bibitem{pdg2010}
K.~Nakamura, {\em et al.} (Particle Data Group), J.\ Phys.\ G {\bf37}, 075021 (2010).



\bibitem{Aerts:1983hy}
  A.~T.~M.~Aerts and C.~B.~Dover,
  Phys.\ Rev.\  D {\bf 28}, 450 (1983).

\bibitem{Iwasaki:1987db}
  Y.~Iwasaki, T.~Yoshie and Y.~Tsuboi,
  Phys.\ Rev.\ Lett.\  {\bf 60}, 1371 (1988).

\bibitem{Stotzer:1997vr}
  R.~W.~Stotzer {\it et al.}  [BNL E836 Collaboration],
  Phys.\ Rev.\ Lett.\  {\bf 78}, 3646 (1997).

\bibitem{Ahn:1996hw}
  J.~K.~Ahn {\it et al.}  [E224 Collaboration],
  Phys.\ Lett.\  B {\bf 378} (1996) 53.

\bibitem{Froemel:2004ea}
  F.~Fr\"oemel, B.~Juli\'{a}-D\'{\i}az and D.~O.~Riska,
  Nucl.\ Phys.\  A {\bf 750}, 337 (2005)
  [arXiv:nucl-th/0410034].

\bibitem{JuliaDiaz:2004rf}
  B.~Juli\'{a}-D\'{\i}az and D.~O.~Riska,
  Nucl.\ Phys.\  A {\bf 755}, 431 (2005)
  [arXiv:nucl-th/0405061].


\bibitem{SprungEtc}
R.~de~Tourreil, B.~Rouben and D.~W.~L.~Sprung, Nucl.\ Phys.\ A {\bf242} 445 (1975);
M.~Lacombe {\em et al.}, Phys.\ Rev.\ C {\bf21} 861 (1980);
R.~Blankenbecler and R.~Sugar, Phys.\ Rev.\ 142 (1966) 1051;
R.~B.~Wiringa, R.~A.~Smith and T.~L.~Ainsworth, Phys.\ Rev.\ C {\bf29} 1207 (1984).

\bibitem{HoukAndLeun}
T.~L.~Houk, Phys.\ Rev.\ C {\bf3} 1886 (1971);
C.~van~der~Leun and C.~Alderliesten, Nucl.\ Phys.\ A {\bf380} 261 (1982).

\bibitem{BerardAndSimon}
G.~G.~Simon, Ch.~Schmitt and V.~H.~Walther, Nucl.\ Phys.\ A 364 (1981) 285;
R.~W.~B\'erard {\em et al.}, Phys.\ Lett.\ B {\bf47} 355 (1973).


\bibitem{Pakhlova:2008vn}
  G.~Pakhlova {\it et al.}  [Belle Collaboration],
  Phys.\ Rev.\ Lett.\  {\bf 101}, 172001 (2008)
  arXiv:0807.4458 [hep-ex].

\bibitem{Aubert:2010yz}
  B.~Aubert {\it et al.}  [BABAR Collaboration],
  Phys.\ Rev.\  D {\bf 82}, 091102 (2010)
  arXiv:1006.2216 [hep-ex].

\bibitem{Collaboration:2011yf}
  N.~S.~f.~Collaboration,
  arXiv:1104.3311 [nucl-ex].








\end{thebibliography}

\section*{APPENDIX}


\subsection{The functions $H_0$, $H_1$, $H_2$ and $H_3$}

The functions $H_0$, $H_1$, $H_2$ and $H_3$ are defined as~\cite{Ding}
\begin{eqnarray}
H_0(\Lambda,m,r) &=& \frac{1}{mr}\left(e^{-mr}-e^{-\Lambda
r}\right)-\frac{\Lambda^2-m^2}{2m\Lambda}e^{-\Lambda r},\NO\\
H_1(\Lambda,m,r) &=& -\frac{1}{mr}(e^{-mr}-e^{-\Lambda r})+
\Lambda\frac{\Lambda^2-m^2}{2m^3}e^{-\Lambda r},\NO\\
H_2\left(\Lambda,m,r\right) &=& \left(1+\frac{1}{mr}
\right)\frac{e^{-mr}}{m^2r^2}-\left(1+\frac{1}{\Lambda r}\right)
\frac{\Lambda}{m}\frac{e^{-\Lambda r}}{m^2r^2}
-\frac{\Lambda^2-m^2}{2m^2}\frac{e^{-\Lambda r}}{mr},\NO\\
H_3\left(\Lambda,m,r\right) &=& \left(1+\frac{3}{mr}+\frac{3}{m^2r^2}\right)\frac{e^{-mr}}{mr}-
\left(1+\frac{3}{\Lambda r}+\frac{3}{\Lambda^2 r^2}\right)\frac{\Lambda^2}{m^2}
\frac{e^{-\Lambda r}}{mr}-\frac{\Lambda^2-m^2}{2m^2}\left(1+\Lambda r\right)\frac{e^{-\Lambda r}}{mr}.
\end{eqnarray}
With Fourier transformations we have
\begin{eqnarray}
\frac{1}{m^2+\bm Q^2} &\rightarrow& \frac{m}{4\pi}H_0(\Lambda,m,r),\NO\\
\frac{\bm Q^2}{m^2+\bm Q^2} &\rightarrow& \frac{m^3}{4\pi}H_1(\Lambda,m,r),\NO\\
\frac{\bm Q}{m^2+\bm Q^2} &\rightarrow& \frac{im^3\bm r}{4\pi}
H_2(\Lambda,m,r),\NO\\
\frac{Q_iQ_j}{m^2+\bm Q^2}&\rightarrow& -\frac{m^3}{12\pi}\left\{H_3(\Lambda,m,r)
\left(3\frac{r_ir_j}{r^2}-\delta_{ij}\right)-H_1(\Lambda,m,r)\delta_{ij}\right\}.
\end{eqnarray}

\subsection{The coupling constants of the heavy baryons and light mesons}

In the quark model we have
\begin{eqnarray}
\GpiXX=0,\;\; \GpiSS=\frac{4}{3}\GpiQQ\frac{m_{\S_c}}{m_q},\;\;
\GpiXXp=\frac{2}{3}\GpiQQ\frac{m_{\Xi^\p_c}}{m_q}, \NO\\ \GetaLL=0,\;\;
\GetaXX=0,\;\;
\GetaSS=\frac{4}{3}\GetaQQ\frac{m_{\S_c}}{m_q},\NO\\ \GetaXXp=-\frac{2}{3}\GetaQQ\frac{m_{\Xi_c^\p}}{m_q},\;\;
\GetaOO=-\frac{8}{3}\GetaQQ\frac{m_{\O_c}}{m_q},\NO\\ \GrhoXX=\GrhoQQ,\;\;
\FrhoXX=-\GrhoQQ,\NO\\ \GrhoSS=2\GrhoQQ,\;\;
\FrhoSS=2\GrhoQQ\left(\frac{2}{3}\frac{m_{\S_c}}{m_q}-1\right),\NO\\ \GrhoXXp=\GrhoQQ,\;\;
\FrhoXXp=\GrhoQQ\left(\frac{2}{3}\frac{m_{\Xi_c^\p}}{m_q}-1\right),\NO\\ \GomLL=2\GomQQ,\;\;
\FomLL=-2\GomQQ,\;\; \GomXX=\GomQQ,\;\;
\FomXX=-\GomQQ,\NO\\ \GomSS=2\GomQQ,\;\;
\FomSS=2\GomQQ\left(\frac{2}{3}\frac{m_{\S_c}}{m_q}-1\right),\NO\\ \GomXXp=\GomQQ,\;\;
\FomXXp=\GomQQ\left(\frac{2}{3}\frac{m_{\Xi_c^\p}}{m_q}-1\right),\NO\\ \GphiXX=\GphiQQ,\;\;
\FphiXX=-\GphiQQ,\NO\\ \GphiXXp=\GphiQQ,\;\; \FphiXXp=\GphiQQ\left(\frac{2}{3}\frac{m_{\Xi_c^\p}}{m_q}-1\right),\NO\\ \GphiOO=2\GphiQQ,\;\;
\FphiOO=2\GphiQQ\left(\frac{2}{3}\frac{m_{\O_c}}{m_q}-1\right),\NO\\ \GsigLL=2\GsigQQ,\;\;
\GsigXX=2\GsigQQ,\NO\\ \GsigSS=2\GsigQQ,\;\; \GsigXXp=2\GsigQQ,\;\;
\GsigOO=2\GsigQQ.
\end{eqnarray}
Because nucleons do not interact directly with the $\phi$ meson in
the quark model, we can not get $\GphiQQ$ in this way. However,
using the $SU(3)$ flavor symmetry, we have
$\GphiQQ=\sqrt{2}\GrhoQQ$. Since $\GrhoQQ$ is related to
$\GrhoNN$, all coupling constants of heavy charmed baryons and
$\phi$ can be expressed in terms of $\GrhoNN$.

\subsection{The dependence of the binding energy on the cutoff}

Finally, we plot the variations of the binding energies with the
cutoff.
\begin{figure}[htb]
	  \hfill
	  \begin{minipage}[t]{0.33\textwidth}
	  \centering
	  \includegraphics[width=0.92\textwidth]{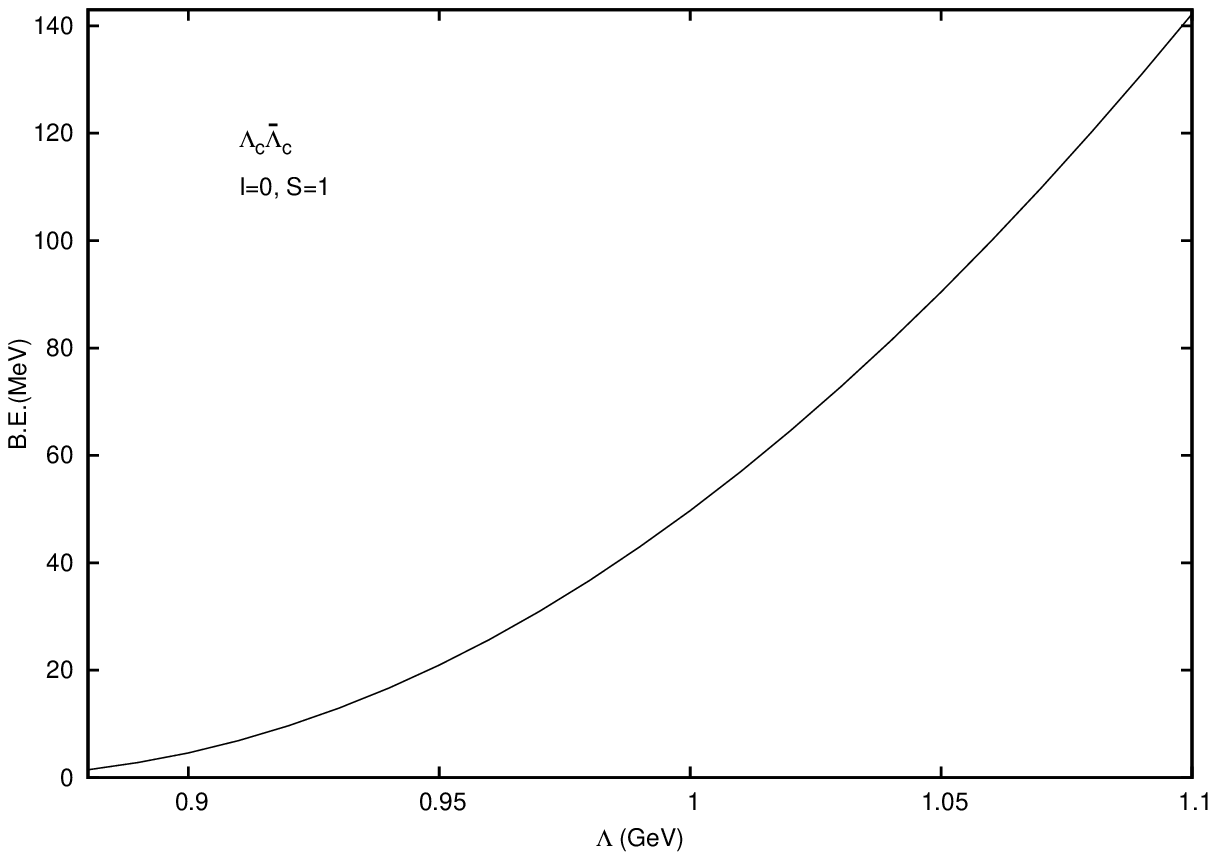}\\ (a)
	  \end{minipage}%
	  \hfill
	  \begin{minipage}[t]{0.33\textwidth}
	  \centering
	  \includegraphics[width=0.92\textwidth]{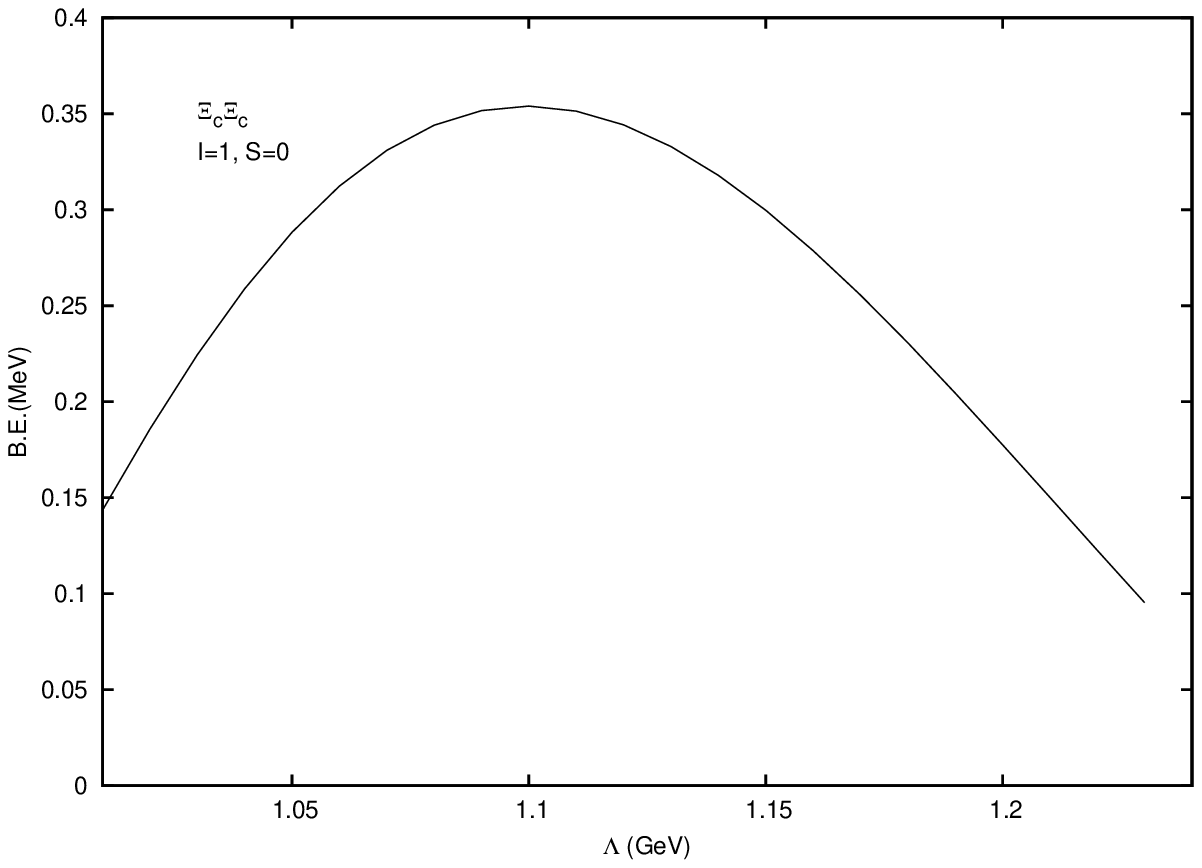}\\ (b)
	  \end{minipage}%
	  \hfill
	  \begin{minipage}[t]{0.33\textwidth}
	  \centering
	  \includegraphics[width=0.92\textwidth]{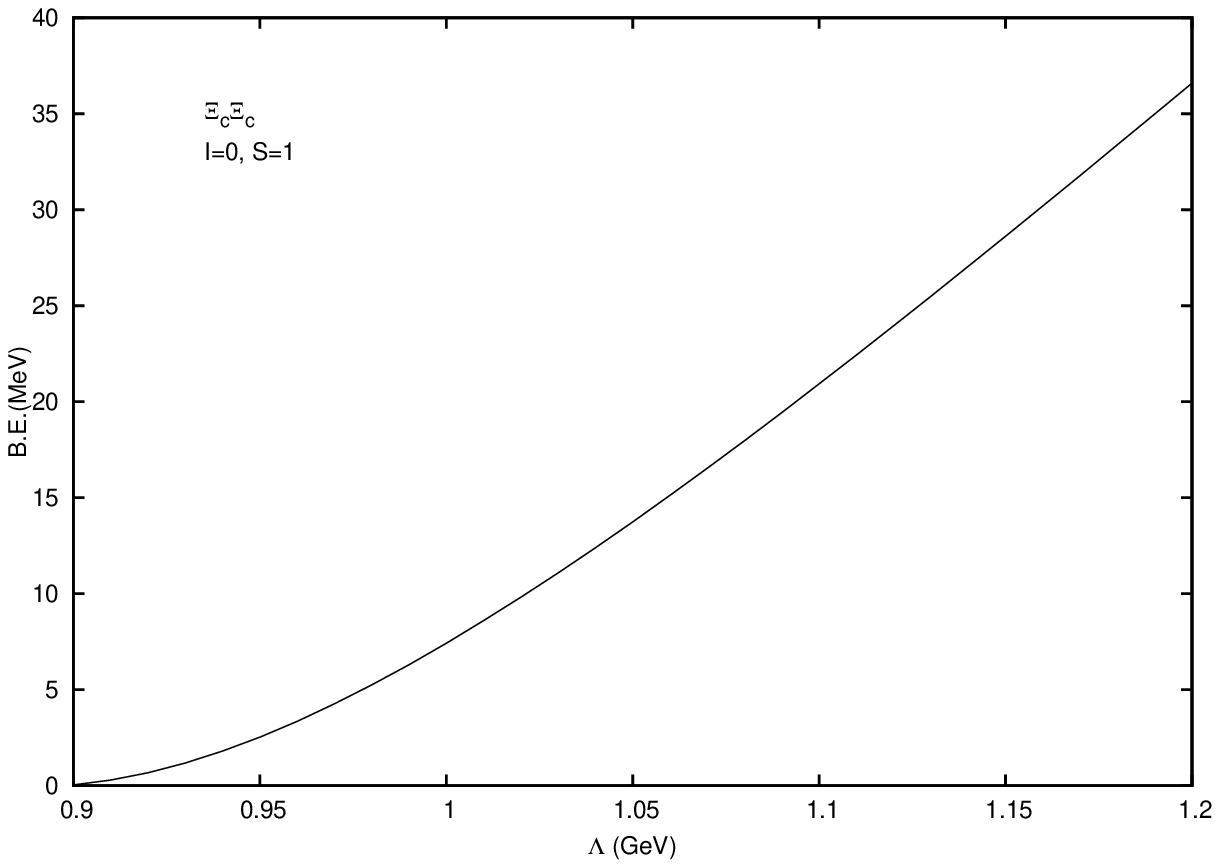}\\ (c)
	  \end{minipage}%
	  \hfill
	  \begin{minipage}[t]{0.33\textwidth}
	  \centering
	  \includegraphics[width=0.9\textwidth]{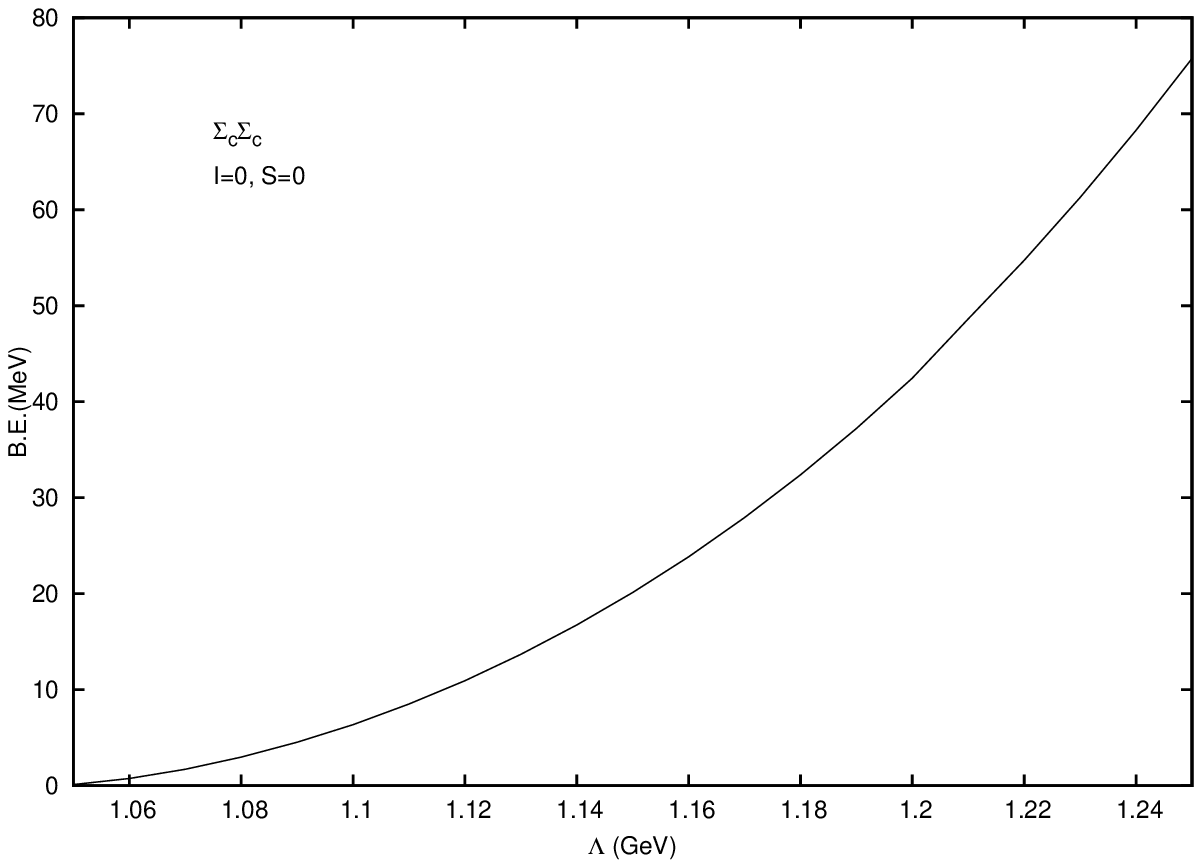}\\ (d)
	  \end{minipage}%
	  \hfill
	  \begin{minipage}[t]{0.33\textwidth}
	  \centering
	  \includegraphics[width=0.9\textwidth]{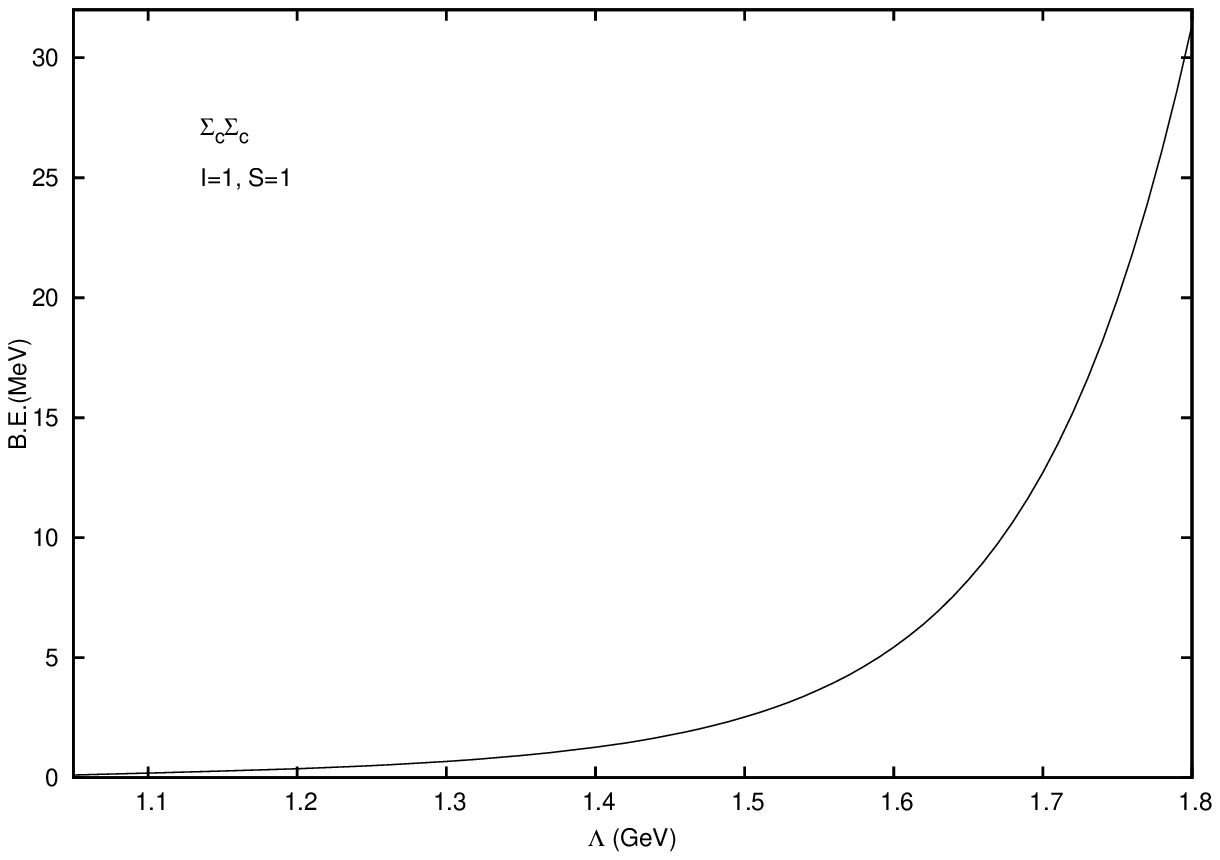}\\ (e)
	  \end{minipage}%
	  \hfill
	  \begin{minipage}[t]{0.33\textwidth}
	  \centering
	  \includegraphics[width=0.9\textwidth]{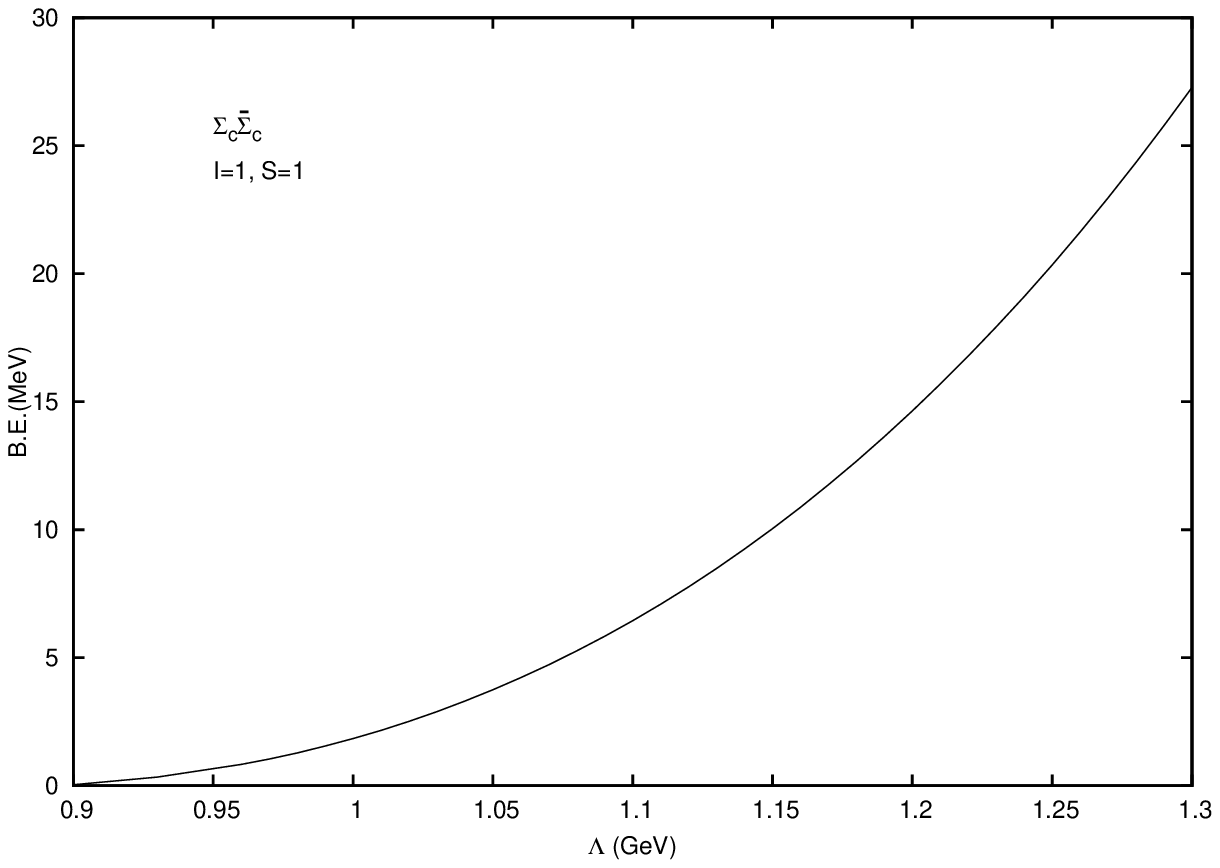}\\ (f)
	  \end{minipage}%
	  \hfill
	  \begin{minipage}[t]{0.33\textwidth}
	  \centering
	  \includegraphics[width=0.9\textwidth]{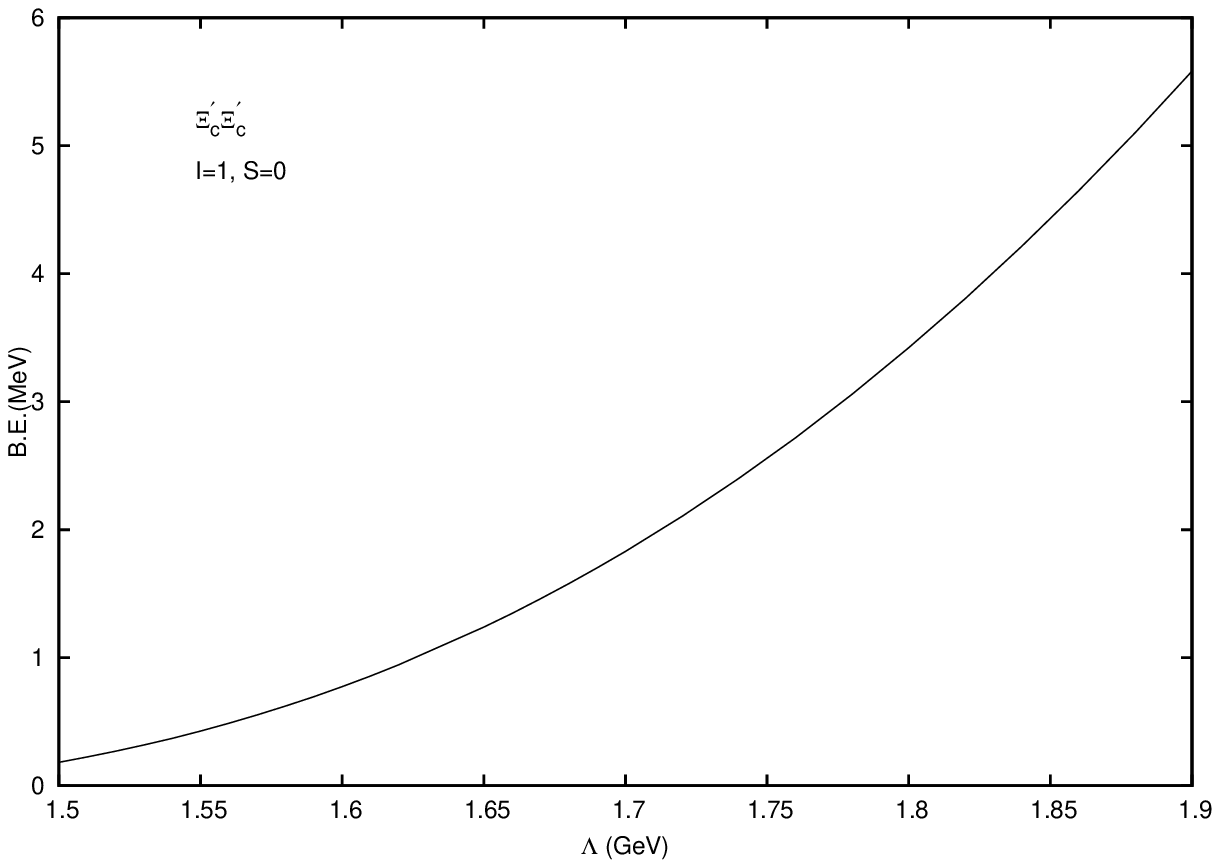}\\ (g)
	  \end{minipage}%
	  \hfill
	  \begin{minipage}[t]{0.33\textwidth}
	  \centering
	  \includegraphics[width=0.9\textwidth]{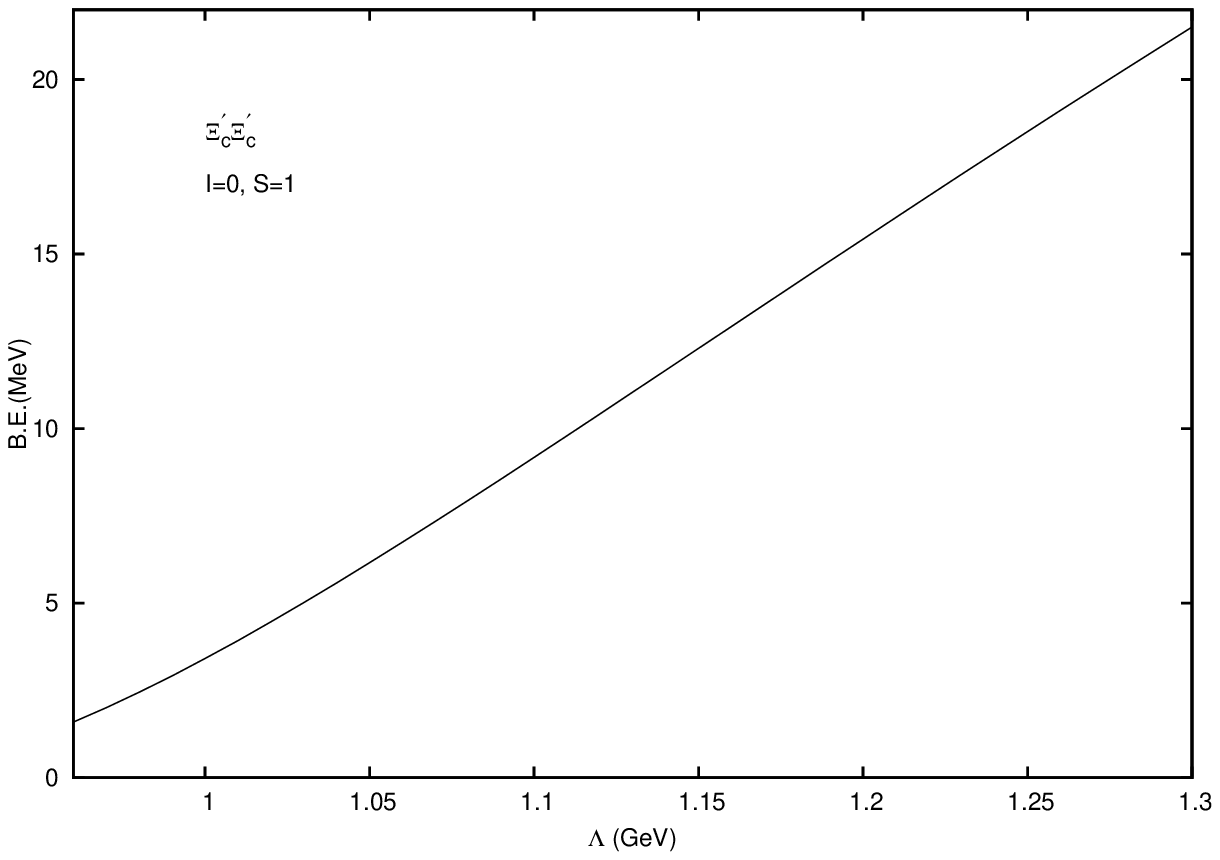}\\ (h)
	  \end{minipage}%
	  \hfill
	  \begin{minipage}[t]{0.33\textwidth}
	  \centering
	  \includegraphics[width=0.9\textwidth]{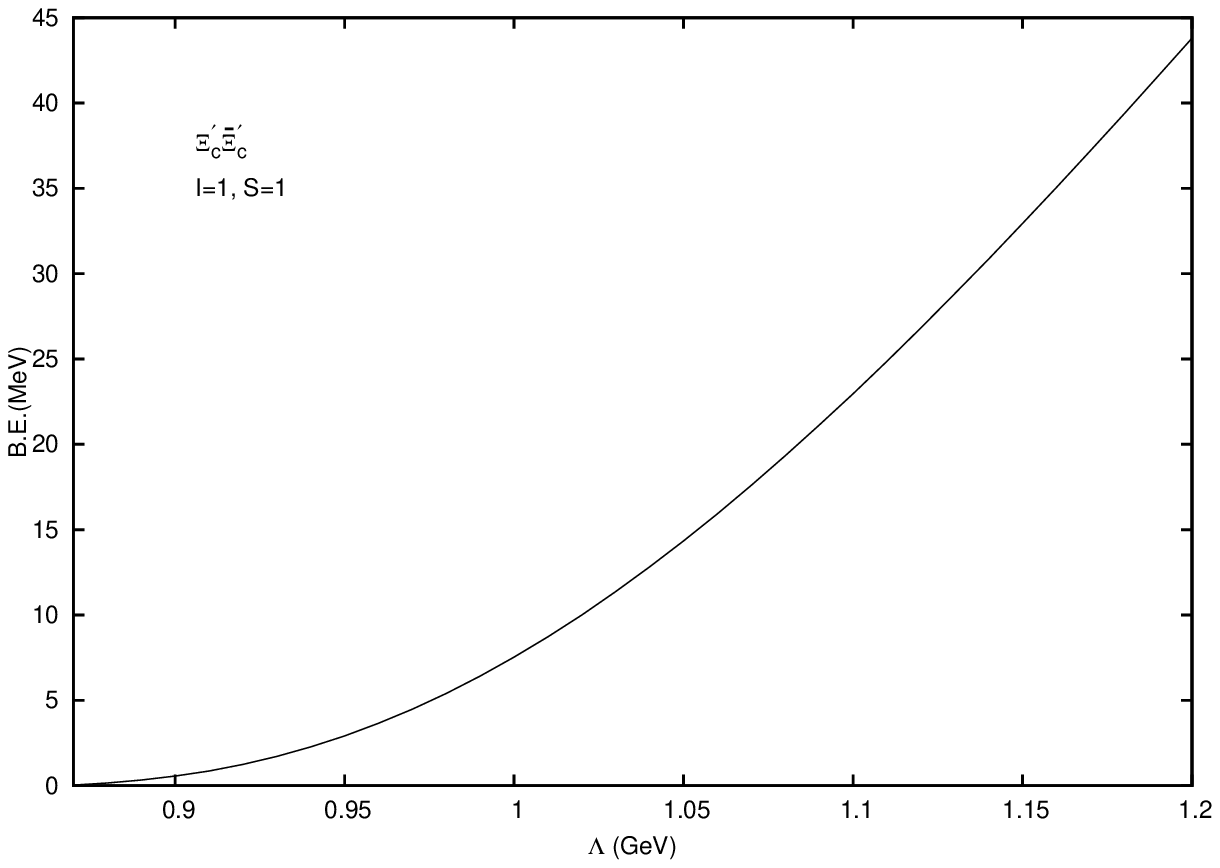}\\ (i)
	  \end{minipage}%
	  \hfill
	  \begin{minipage}[t]{0.33\textwidth}
	  \centering
	  \includegraphics[width=0.9\textwidth]{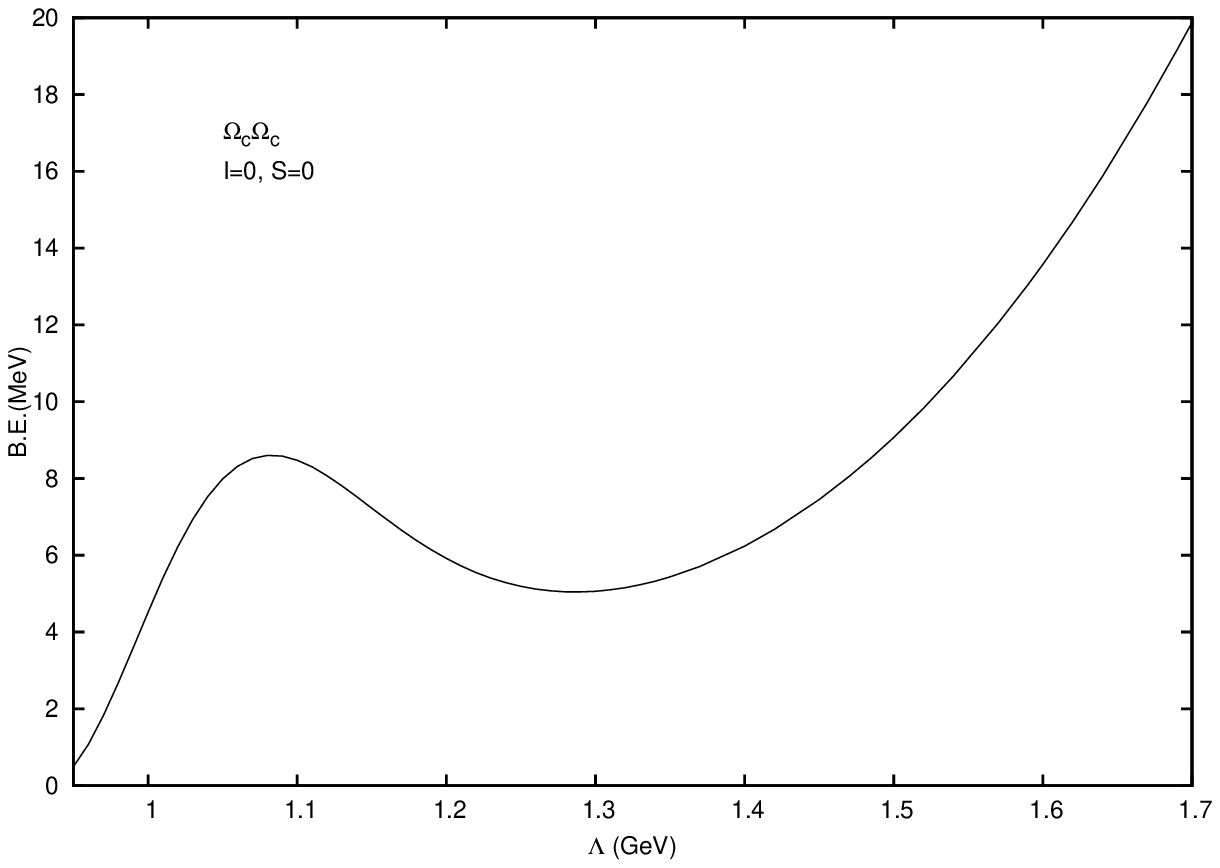}\\ (j)
	  \end{minipage}%
	  \hfill
	  \begin{minipage}[t]{0.33\textwidth}
	  \centering
	  \includegraphics[width=0.9\textwidth]{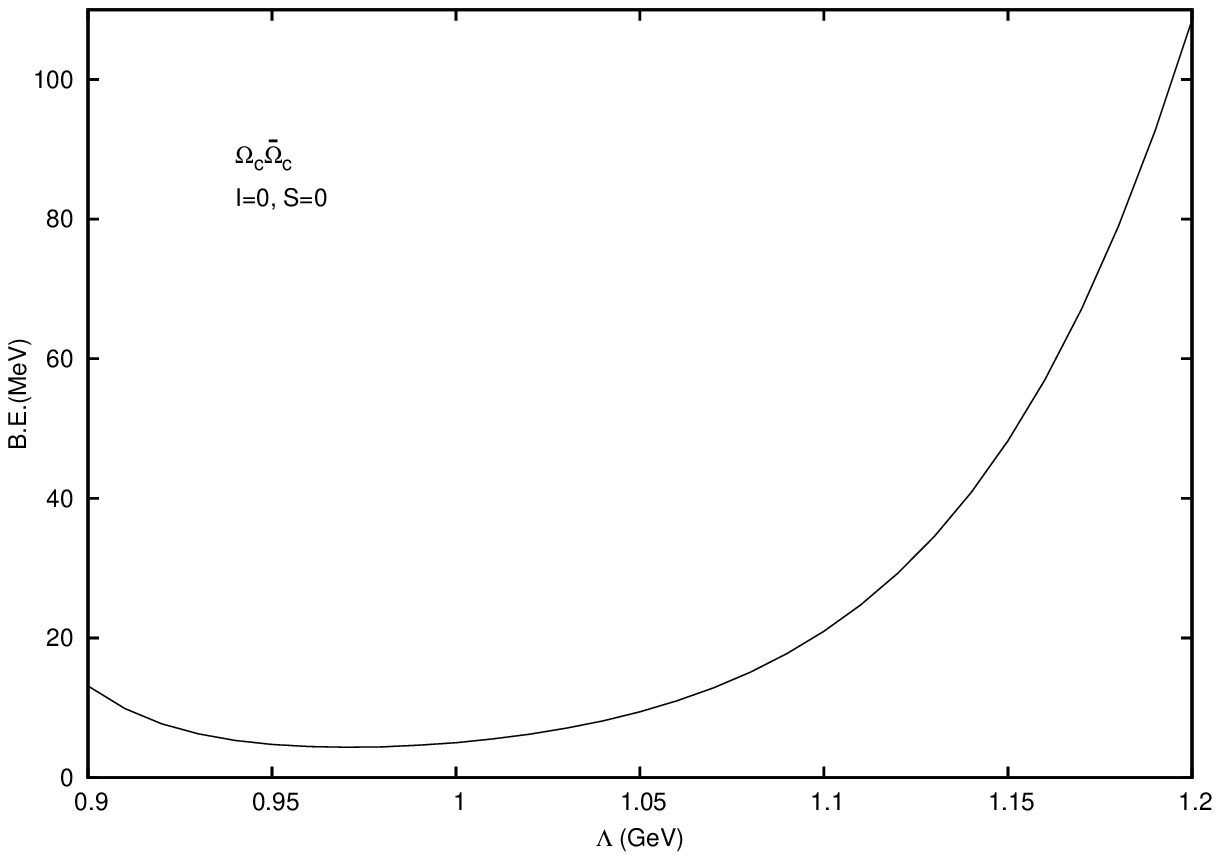}\\ (k)
	  \end{minipage}%
	  \hfill
	  \begin{minipage}[t]{0.33\textwidth}
	  \centering
	  \includegraphics[width=0.9\textwidth]{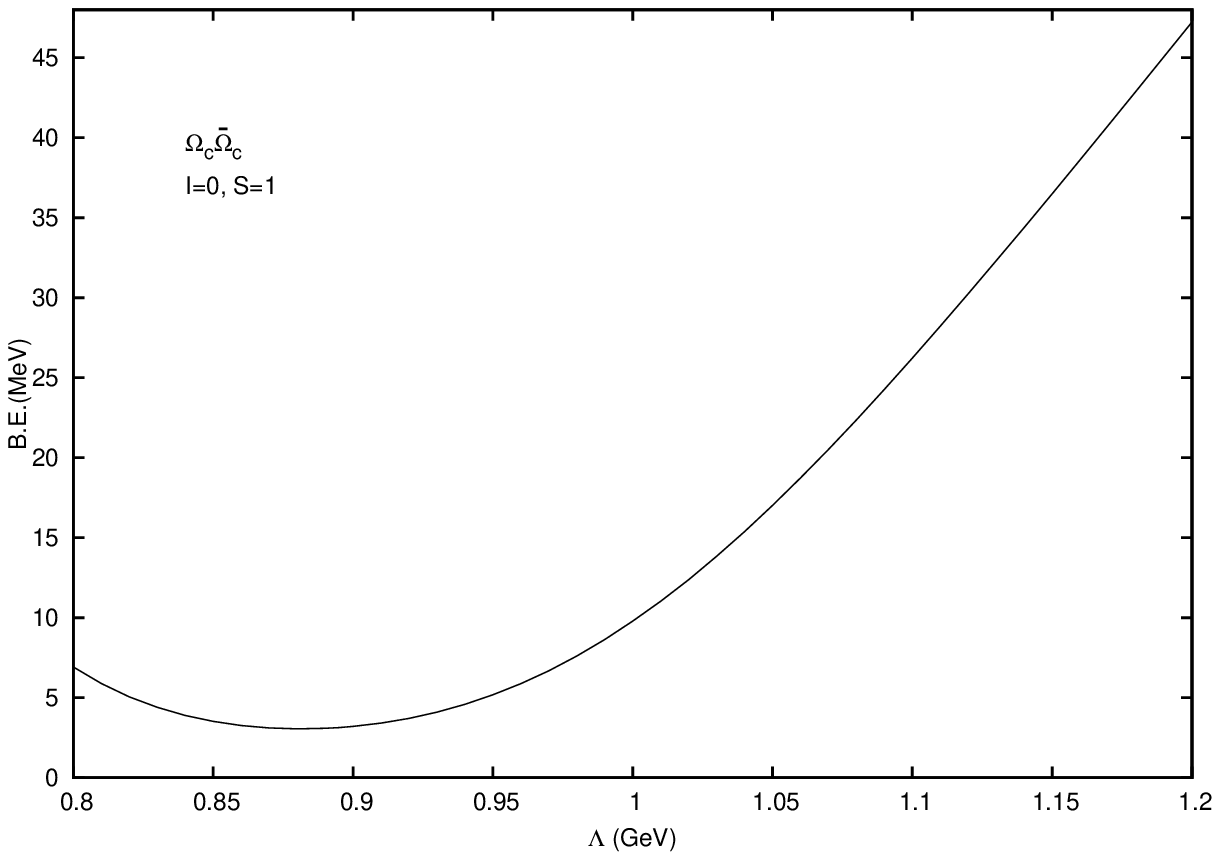}\\ (l)
	  \end{minipage}%
	  \hspace*{\fill}
  \caption{Dependence of the binding energy on the cutoff.
In Figs.(f) and (i), only one-pion contributions are included.}
  \label{Fig:EbvsL} 
\end{figure}
\end{document}